\documentclass[graybox]{svmult}
\usepackage{mathptmx}       % selects Times Roman as basic font
\usepackage{helvet}         % selects Helvetica as sans-serif font
\usepackage{courier}        % selects Courier as typewriter font
\usepackage{type1cm}        % activate if the above 3 fonts are
\usepackage{makeidx}         % allows index generation
\usepackage{graphicx}        % standard LaTeX graphics tool
\usepackage{multicol}        % used for the two-column index
\usepackage[bottom]{footmisc}% places footnotes at page bottom

\usepackage{amsmath}
\usepackage{amssymb}

\makeindex             % used for the subject index
\begin{document}

\title*{Classical Yang-Mills Black Hole Hair in anti-de Sitter Space}
\author{Elizabeth Winstanley}
\institute{Elizabeth Winstanley \at School of Mathematics and Statistics,
The University of Sheffield, Hicks Building, Hounsfield Road, Sheffield S3 7RH, United Kingdom.
\email{E.Winstanley@sheffield.ac.uk}}

\maketitle

\abstract{The properties of hairy black holes in Einstein-Yang-Mills (EYM) theory are reviewed,
focusing on spherically symmetric solutions.
In particular, in asymptotically anti-de Sitter space (adS) stable black hole hair is
known to exist for ${\mathfrak {su}}(2)$ EYM.
We review recent work in which it is shown that stable hair also exists in ${\mathfrak {su}}(N)$ EYM
for arbitrary $N$, so that there is no upper limit on how much stable hair a black hole in
adS can possess.}

\section{Introduction}
\label{win:sec:intro}

We begin by very briefly reviewing the ``no-hair'' conjecture and motivating the study of hairy black holes.

\subsection{The ``no-hair'' conjecture}
\label{win:sec:nohair}

The black hole ``no-hair'' conjecture \cite{win:Wheeler1} states that (see, for example,
\cite{win:Chrusciel1,win:Chrusciel2,win:Heusler1,win:Heusler3,win:Heusler2,win:Mazur1} for
detailed reviews and comprehensive lists of references):
\begin{svgraybox}
All stationary, asymptotically flat, four-dimensional black hole equilibrium solutions of the Einstein
equations in vacuum or with an electromagnetic field are characterized by their mass,
angular momentum, and (electric or magnetic) charge.
\end{svgraybox}
According to the no-hair conjecture,
black holes are therefore extraordinarily simple objects, whose geometry
(exterior to the event horizon) is a member of the Kerr-Newman
family and completely determined by just three quantities (mass, angular momentum and charge).
Furthermore, these quantities are {\em {global charges}} which can (at least in principle) be measured
at infinity, far from the black hole event horizon.
If a black hole is formed by the gravitational collapse of a dying star, the initial star will be a highly
complex object described by many different parameters.
The final, equilibrium, black hole is, by contrast, rather simple and described by a very small number of
quantities.
During the process of the formation of a black hole, an enormous amount of (classical) information
about the star which collapsed has therefore been lost.
Similarly, if a complicated object is thrown down a black hole event horizon, once the system settles
down, the only changes in the final state will be changes in the total mass, total angular momentum
and total charge.
Advances in astrometry \cite{win:Will1} and future gravitational wave detectors \cite{win:Barack1} may even be
able to probe the validity of the ``no-hair'' conjecture for astrophysical black holes by verifying that the mass,
angular momentum and quadrupole moment $Q_{2}$ of the black hole satisfy the relation $Q_{2}=J^{2}/M$ which holds for
Kerr black holes.

The ``no-hair'' conjecture, stated above, has been proved by means of much complicated and beautiful mathematics
(as reviewed in, for example, \cite{win:Chrusciel1,win:Chrusciel2,win:Heusler1,win:Heusler3,win:Heusler2,win:Mazur1}),
subject to the assumptions of stationarity,
asymptotic flatness, four-dimensional space-time and the electrovac Einstein equations.
It is perhaps unsurprising that if one or more of these assumptions is relaxed, then the
conjecture does not necessarily hold.
For example, if a negative cosmological constant is included, so that the space-time is no longer asymptotically
flat but instead approaches anti-de Sitter (adS) space at infinity, then the event horizon
of the black hole is not necessarily spherical, giving rise to ``topological'' black holes
(see, for example, \cite{win:Birmingham1,win:Galloway1,win:Klemm1,win:Lemos1,win:Mann1,win:Vanzo1}).
More recently, the discovery of ``black ring'' solutions in five space-time dimensions
(\cite{win:Emparan1}, see \cite{win:Emparan2} for a recent review)
and the even more complicated ``black Saturn'' \cite{win:Elvang1} solutions
indicates that Einstein-Maxwell theory has a rich space of black solutions in higher dimensions,
which are not given in terms of the Myers-Perry \cite{win:Myers1} metric
(which is the generalization of the Kerr-Newman geometry to higher dimensions).

\subsection{Hairy black holes}
\label{win:sec:hairyBH}

In this article we consider what happens when the other condition in the ``no-hair'' conjecture,
namely that the Einstein equations involve electrovac matter only, is relaxed.
The ``generalized'' version of the no-hair conjecture \cite{win:Heusler2} states
that all stationary black hole solutions of the Einstein equations with any type of
self-gravitating matter field are determined uniquely by their mass, angular momentum
and a set of global charges.
Even in asymptotically flat space, this conjecture does not hold, even for the simplest type
of self-gravitating matter, a scalar field.
The first such counter-example is the famous BBMB black hole
\cite{win:Bekenstein1,win:Bekenstein2,win:Bocharova} which has the same metric as the extremal
Reissner-Nordstr\"om black hole but possesses a conformally coupled scalar field.
However, this solution is controversial due to the divergence of the scalar field on the event horizon
\cite{win:Sudarsky1} and is also highly unstable \cite{win:Bronnikov1}.
Therefore, in some ways the first ``hairy'' black hole is considered to be the Gibbons solution
\cite{win:Gibbons1}, which describes a Reissner-Nordstr\"om black hole with a non-trivial dilaton field.
While there are many results which rule out scalar field hair in quite general models,
particularly in asymptotically flat space-times (see, for example, \cite{win:Bekenstein3} for a review),
in recent years many other examples of black holes with non-trivial scalar field hair have been found.
For example, minimally coupled scalar field hair has been found when the cosmological constant is
positive \cite{win:Torii1} or negative \cite{win:Torii2} and non-minimally coupled scalar field hair
has also been considered (see, for example, \cite{win:ew2,win:ew4} and references therein).

In this short review, we will focus on another particular matter model, Einstein-Yang-Mills theory (EYM), where the
matter is described by a non-Abelian (Yang-Mills) gauge field.
It is now well-known that this theory possesses ``hairy'' black hole solutions,
whose metric is not a member of the Kerr-Newman family (see \cite{win:Volkov1} for a
detailed review).
Furthermore, unlike the Kerr-Newman black holes, the geometry exterior to the event horizon
is not determined uniquely by global charges measureable at infinity,
although only a small number of parameters are required in order to describe the metric
and matter field
(see section \ref{win:sec:af} for further details).
All of the asymptotically flat black hole solutions of pure EYM theory
discovered to date are unstable \cite{win:Brodbeck1}
(however, there are examples of asymptotically flat, stable hairy black holes in variants of the
EYM action, such as Einstein-Skyrme \cite{win:Bizon2,win:Droz1,win:Droz2,win:Heusler4},
Einstein-non-Abelian-Proca \cite{win:Greene1,win:Maeda1,win:Torii5,win:Tamaki1,win:Torii4}
and Einstein-Yang-Mills-Higgs \cite{win:Aichelburg1} theories).
This means that, while the ``letter'' of the no-hair theorem is violated in this case (as there
exist solutions which are not described by the Kerr-Newman metric), its ``spirit'' is intact,
as stable equilibrium black holes remain simple objects, described by a few parameters if not
exactly of the Kerr-Newman form (see \cite{win:Bizon6} for a related discussion along these lines).

The situation is radically different if one considers EYM solutions in asymptotically adS
space, rather than asymptotically flat space.
For ${\mathfrak {su}}(2)$ EYM, at least some black hole solutions with hair are
stable \cite{win:Bjoraker1,win:Bjoraker2,win:ew1}.
These stable black holes require one new parameter (see section \ref{win:sec:su2adS}) to
completely describe the geometry exterior to their event horizons.
Therefore, one might still argue that the true ``spirit'' of the ``no-hair'' conjecture remains intact,
that stable equilibrium black holes are comparatively simple objects, described by just a few parameters.

One is therefore led to a natural question:  are there hairy black hole solutions in adS which
require an infinite number of parameters to fully describe the geometry and matter exterior to the
event horizon?
In other words, is there a limit to how much hair a black hole in adS can be given?
This is the question we will be seeking to address in this article.

\subsection{Scope of this article}
\label{win:sec:note}

The subject of hairy black holes in EYM theory and its variants is very active,
with many new solutions appearing each year.
The review \cite{win:Volkov1}, written in 1998, is very detailed and thorough, and contains
a comprehensive list of references to solutions known at that time.
We have therefore not sought to be complete in our references prior to that date, and have, instead,
chosen to highlight a few solutions (the selection being undoubtedly personal).
Even considering just work after 1998, we have been unable to do justice to the
huge body of work in this area (for example, the seminal paper \cite{win:Bartnik2} has 172
arXiv citations between 1999 and the time of writing), and have instead chosen some examples
of solutions.
As well as \cite{win:Volkov1}, reviews of various aspects of solitons and black holes in EYM can be found in
\cite{win:Bizon6,win:Galtsov2,win:Gibbons2,win:Straumann4,win:Straumann3,win:Volkov7}.

The outline of this article is as follows.
In section \ref{win:sec:suN} we will outline ${\mathfrak {su}}(N)$ EYM theory,
including our ansatz for the gauge field and the form of the field equations.
We will then, in section \ref{win:sec:af}, briefly review some of the properties of the well-known
asymptotically flat solutions of this theory.
Our main focus in this article are asymptotically adS black holes, and we begin
our discussion of these in section \ref{win:sec:su2adS} by reviewing the key features
of the ${\mathfrak {su}}(2)$ EYM black holes in adS, before moving on to describe
very recent work on ${\mathfrak {su}}(N)$, asymptotically adS, EYM black holes in section \ref{win:sec:suNadS}.
Our conclusions are presented in section \ref{win:sec:conc}.
Throughout this article the metric has signature $(-,+,+,+)$ and we use units in which $4\pi G=c=1$.

\section{${\mathfrak {su}}(N)$ Einstein-Yang-Mills theory}
\label{win:sec:suN}

In this section we gather together all the formalism and field equations we shall require
for our later study of black hole solutions.

\subsection{Ansatz, field equations and boundary conditions}
\label{win:sec:ansatz}

In this article we shall be interested in four-dimensional
${\mathfrak {su}}(N)$ EYM theory with a cosmological constant, described by the
following action, given in suitable units:
\begin{equation}
S_{\mathrm {EYM}} = \frac {1}{2} \int \D ^{4}x {\sqrt {-g}} \left[
R - 2\Lambda - {\mathrm {Tr}} \, F_{\mu \nu }F ^{\mu \nu }
\right] ,
\label{win:eq:action}
\end{equation}
where $R$ is the Ricci scalar of the geometry and $\Lambda $ the cosmological constant.
Here we have chosen the simplest type of EYM-like theory, many variants have been studied in the literature
(see, for example, \cite{win:Volkov1} for a selection of examples).

Varying the action (\ref{win:eq:action}) gives the field equations
\begin{eqnarray}
T_{\mu \nu } & = & R_{\mu \nu } - \frac {1}{2} R g_{\mu \nu } + \Lambda g_{\mu \nu };
\nonumber \\
0 & = & D_{\mu } F_{\nu }{}^{\mu } = \nabla _{\mu } F_{\nu }{}^{\mu }
+ \left[ A_{\mu }, F_{\nu }{}^{\mu } \right] ;
\end{eqnarray}
where the YM stress-energy tensor is
\begin{equation}
T_{\mu \nu } = \, {\mbox {Tr}} \, F_{\mu \lambda } F_{\nu }{}^{\lambda } - \frac {1}{4} g_{\mu \nu }
{\mbox {Tr}} \, F_{\lambda \sigma} F^{\lambda \sigma } .
\label{win:eq:Tmunu}
\end{equation}
In this article we consider only static, spherically symmetric black hole geometries, with metric given,
in standard Schwarzschild-like co-ordinates, as
\begin{equation}
ds^{2} = - \mu S^{2} \, dt^{2} + \mu ^{-1} \, dr^{2} +
r^{2} \, d\theta ^{2} + r^{2} \sin ^{2} \theta \, d\phi ^{2} ,
\label{win:eq:metric}
\end{equation}
where the metric functions $\mu $ and $S$ depend on the radial co-ordinate $r$ only.
In the presence of a negative cosmological constant $\Lambda <0$, we write the metric function $\mu $ as
\begin{equation}
\mu (r) = 1 - \frac {2m(r)}{r} - \frac {\Lambda r^{2}}{3}.
\label{win:eq:mu}
\end{equation}

The most general, spherically symmetric, ansatz for the ${\mathfrak {su}}(N)$ gauge potential is \cite{win:Kunzle1}:
\begin{equation}
A  =
{\cal {A}} \, dt + {\cal {B}} \, dr +
\frac {1}{2} \left( C - C^{H} \right) \, d\theta
- \frac {i}{2} \left[ \left(
C + C^{H} \right) \sin \theta + D \cos \theta \right] \, d\phi ,
\label{win:eq:gaugepot}
\end{equation}
where ${\cal {A}}$, ${\cal {B}}$, $C$ and $D$ are all $\left( N \times N \right) $ matrices and
$C^{H}$ is the Hermitian conjugate of $C$.
The matrices ${\cal {A}}$ and ${\cal {B}}$ are purely imaginary, diagonal, traceless and depend only
on the radial co-ordinate $r$.
The matrix $C$ is upper-triangular, with non-zero entries only immediately above the diagonal:
\begin{equation}
C_{j,j+1}=\omega_j (r) e^{i\gamma _{j}(r)},
\end{equation}
for $j=1,\ldots,N-1$.
In addition, $D$ is a constant matrix:
\begin{equation}
D=\mbox{Diag}\left(N-1,N-3,\ldots,-N+3,-N+1\right) .
\label{win:eq:matrixD}
\end{equation}
Here we are primarily interested only in purely magnetic solutions, so we set ${\cal {A}} \equiv 0$.
We may also take ${\cal {B}}\equiv 0$ by a choice of gauge \cite{win:Kunzle1}.
From now on we will assume that all the $\omega _{j}(r)$ are non-zero
(see, for example, \cite{win:Galtsov1,win:Kleihaus1,win:Kleihaus2,win:Kleihaus3}
for the possibilities in asymptotically flat space if this
assumption does not hold).
In this case one of the Yang-Mills equations becomes \cite{win:Kunzle1}
\begin{equation}
\gamma _{j} = 0 \qquad \forall j=1,\ldots , N-1.
\end{equation}
Our ansatz for the Yang-Mills potential therefore reduces to
\begin{equation}
A = \frac {1}{2} \left( C - C^{H} \right) \, d\theta - \frac {i}{2} \left[ \left(
C + C^{H} \right) \sin \theta + D \cos \theta \right] \, d\phi ,
\label{win:eq:gaugepotsimple}
\end{equation}
where the only non-zero entries of the matrix $C$ are
\begin{equation}
 C _{j,j+1} = \omega _{j}(r).
\end{equation}
The gauge field is therefore described by the $N-1$ functions $\omega _{j}(r)$.
We comment that our ansatz (\ref{win:eq:gaugepotsimple}) is by no means the only possible choice in
${\mathfrak {su}}(N)$ EYM.
Techniques for finding {\em {all}} spherically symmetric ${\mathfrak {su}}(N)$ gauge potentials
can be found in \cite{win:Bartnik1}, where all irreducible models are explicitly listed for $N\le 6$.

With the ansatz (\ref{win:eq:gaugepotsimple}), there are $N-1$ non-trivial Yang-Mills equations
for the $N-1$ functions $\omega _{j}$:
\begin{equation}
r^2\mu\omega''_{j}+\left(2m-2r^3 p_{\theta}-\frac{2\Lambda r^3}{3}\right)\omega'_{j}+W_j\omega_j=0
\label{win:eq:YMe}
\end{equation}
for $j=1,\ldots,N-1$, where a prime $'$ denotes $\D /\D r $,
\begin{eqnarray}
p_{\theta}&=&
\frac{1}{4r^4}\sum^N_{j=1}\left[\left(\omega^2_j-\omega^2_{j-1}-N-1+2j\right)^2\right],
\label{win:eq:ptheta}
\\
W_j&=&
1-\omega^2_j+\frac{1}{2}\left(\omega^2_{j-1}+\omega^2_{j+1}\right),
\end{eqnarray}
and $\omega_0=\omega_N=0$.
The Einstein equations take the form
\begin{equation}
m' =
\mu G+r^2p_{\theta},
\qquad
\frac{S'}{S}=\frac{2G}{r},
\label{win:eq:Ee}
\end{equation}
where
\begin{equation}
G=\sum^{N-1}_{j=1}\omega_j'^2.
\label{win:eq:Gdef}
\end{equation}
Altogether, then, we have $N+1$ ordinary differential equations for the $N+1$ unknown functions $m(r)$, $S(r)$
and $\omega _{j}(r)$.
The field equations (\ref{win:eq:YMe},\ref{win:eq:Ee}) are invariant under the transformation
\begin{equation}
\omega _{j} (r) \rightarrow -\omega _{j} (r)
\label{win:eq:omegaswap}
\end{equation}
for each $j$ independently, and also under the substitution:
\begin{equation}
j \rightarrow N - j.
\label{win:eq:Nswap}
\end{equation}

We are interested in black hole solutions of the field equations (\ref{win:eq:YMe},\ref{win:eq:Ee}).
We assume there is a regular, non-extremal, black hole event horizon at $r=r_{h}$, where $\mu (r)$ has
a single zero.
This fixes the value of $m(r_{h})$ to be:
\begin{equation}
2m( r_{h} ) = r_{h} - \frac {\Lambda r_{h}^{3}}{3}.
\end{equation}
However, the field equations (\ref{win:eq:YMe},\ref{win:eq:Ee}) are singular at
the black hole event horizon $r=r_{h}$ and at infinity $r\rightarrow \infty $.
We therefore need to impose boundary conditions on the field variables $m(r)$, $S(r)$ and $\omega _{j}(r)$
at these singular points.
When the cosmological constant $\Lambda $ is zero, local existence of solutions of the field equations
in neighbourhoods of these singular points has been rigorously proved
\cite{win:Kunzle2,win:Oliynyk2}.
This proof can be extended to the case when the cosmological constant is negative \cite{win:Baxter2,win:Baxter1}.

We assume that the field variables $\omega _{j}(r)$, $m(r)$ and $S(r)$  have
regular Taylor series expansions about $r=r_{h}$:
\begin{eqnarray}
m(r) & = & m (r_{h}) + m' (r_{h}) \left( r - r_{h} \right)
+ O \left( r- r_{h} \right) ^{2} ;
\nonumber \\
\omega _{j} (r) & = & \omega _{j}(r_{h}) + \omega _{j}' (r_{h}) \left( r - r_{h} \right)
+ O \left( r -r_{h} \right) ^{2};
\nonumber \\
S(r) & = & S(r_{h}) + S'(r_{h}) \left( r-r_{h} \right)
+ O\left( r - r_{h} \right) .
\label{win:eq:horizon}
\end{eqnarray}
Setting $\mu (r_{h})=0$ in the Yang-Mills equations (\ref{win:eq:YMe}) fixes the derivatives of the
gauge field functions at the horizon:
\begin{equation}
\omega _{j} ' (r_{h}) = - \frac {W_{j}(r_{h})\omega _{j}(r_{h})}{2m(r_{h}) - 2r_{h}^{3} p_{\theta } (r_{h})
-\frac {2\Lambda r_{h}^{3}}{3}}.
\end{equation}
Therefore the expansions (\ref{win:eq:horizon}) are determined by
the $N+1$ quantities $\omega _{j}(r_{h})$, $r_{h}$, $S(r_{h})$ for fixed
cosmological constant $\Lambda $.
For the event horizon to be non-extremal, it must be the case that
\begin{equation}
2m'(r_{h}) = 2r_{h}^{2} p_{\theta} (r_{h}) < 1- \Lambda r_{h}^{2},
\label{win:eq:constraint}
\end{equation}
which weakly constrains the possible values of the gauge field functions $\omega _{j}(r_{h}) $
at the event horizon.
Since the field equations (\ref{win:eq:YMe},\ref{win:eq:Ee}) are invariant under the transformation
(\ref{win:eq:omegaswap}), we may consider
$\omega _{j}(r_{h}) >0$ without loss of generality.

At infinity, we require that the field variables $\omega _{j}(r)$, $m(r)$ and $S(r)$ converge to constant values as
$r\rightarrow \infty $, and have regular Taylor series expansions in $r^{-1}$ near infinity:
\begin{equation}
m(r)   =   M + O \left( r^{-1} \right) ;
\qquad
S(r) = 1 + O\left( r^{-1} \right) ;
\qquad
\omega _{j}(r)  =  \omega _{j,\infty } + O \left( r^{-1} \right) .
\label{win:eq:infinity}
\end{equation}
If the space-time is asymptotically flat, with $\Lambda =0$, then the values of $\omega _{j,\infty }$
are constrained to be
\begin{equation}
\omega _{j,\infty } = \pm {\sqrt {j(N-j)}} .
\end{equation}
This condition means that the asymptotically flat black holes have no magnetic charge at infinity, or,
in other words, these solutions have no global magnetic charge.
Therefore, at infinity, they are indistinguishable from Schwarzschild black holes.
However, if the cosmological constant is non-zero, so that the geometry approaches (a)dS at infinity,
then there are no {\it {a priori}} constraints on the values of $\omega _{j,\infty }$.
In general, therefore, the (a)dS black holes will be magnetically charged.
It should be noted that the boundary conditions in the case when the cosmological constant $\Lambda $ is positive
are more complex, as there is a cosmological horizon between the event horizon and infinity.

\subsection{Some ``trivial'' solutions}
\label{win:sec:trivial}

Although the field equations (\ref{win:eq:YMe},\ref{win:eq:Ee}) are highly non-linear and
rather complicated, they do have some trivial solutions which can easily be written down:
%\begin{svgraybox}
\begin{description}
\item[{\em {Schwarzschild(-(a)dS)}}]
Setting
\begin{equation}
\omega _{j}(r) \equiv \pm {\sqrt {j(N-j)}}
\label{win:eq:SadS}
\end{equation}
for all $j$ gives the Schwarzschild(-(a)dS) black hole
with
\begin{equation}
m(r) =M= {\mbox {constant}}
\end{equation}
We note that, by setting $M=0$, pure Minkowski ($\Lambda =0$) or (a)dS ($\Lambda \neq 0$) space is also a solution.
\item[{\em {Reissner-Nordstr\"om(-(a)dS)}}]
Setting
\begin{equation}
\omega _{j}(r) \equiv 0
\label{win:eq:RNadS}
\end{equation}
for all $j$ gives the Reissner-Nordstr\"om(-(a)dS) black hole
with metric function
\begin{equation}
\mu (r) = 1 - \frac {2M}{r} + \frac {Q^{2}}{r^{2}} - \frac {\Lambda r^{2}}{3},
\end{equation}
where the magnetic charge $Q$ is fixed by
\begin{equation}
Q^{2} = \frac {1}{6} N \left( N+1 \right) \left( N-1 \right) .
\end{equation}
\item[{\em {Embedded ${\mathfrak {su}}(2)$ solutions}}]
%\end{svgraybox}
For our later numerical and analytic work, an additional special class of solutions
turns out to be extremely useful.
We begin by setting
\begin{equation}
\omega _{j}(r) =\pm  {\sqrt {j(N-j)}} \, \omega (r) \qquad \forall j=1,\ldots ,N-1,
\label{win:eq:embeddedsu2}
\end{equation}
then follow \cite{win:Kunzle2} and define
\begin{equation}
\lambda _{N} = {\sqrt {\frac {1}{6} N \left( N -1 \right) \left( N +1  \right) }},
\label{win:eq:lambdaN}
\end{equation}
and then rescale the field variables as follows:
\begin{eqnarray}
R & = &  \lambda _{N}^{-1} r; \qquad
{\tilde {\Lambda }} = \lambda _{N}^{2}\Lambda ; \qquad
{\tilde {m}}(R)  =  \lambda _{N}^{-1} m(r);
\nonumber \\
{\tilde {S}}(R) & =  & S(r); \qquad
{\tilde {\omega }}(R)  =  \omega (r).
\label{win:eq:scaling}
\end{eqnarray}
Note that we rescale the cosmological constant $\Lambda $ (this is not necessary in \cite{win:Kunzle2} as
there $\Lambda = 0$).
The field equations satisfied by ${\tilde {m}}(R)$, ${\tilde {S}}(R)$ and ${\tilde {\omega }}(R)$ are then
\begin{eqnarray}
\frac {d{\tilde {m}}}{dR} & = &
 \mu {\tilde {G}} + R^{2} {\tilde {p}}_{\theta }  ; \qquad
\nonumber \\
\frac {1}{{\tilde {S}}} \frac {d{\tilde {S}}}{dR} & = &
- \frac {2{\tilde {G}}}{R} ;
\nonumber \\
0 & = & R^{2} \mu \frac {d^{2}{\tilde {\omega }}}{dR^{2}} +
\left[ 2{\tilde {m}} - 2R^{3} {\tilde {p}}_{\theta } - \frac {2{\tilde {\Lambda }}R^{3}}{3}
\right] \frac {d{\tilde {\omega }}}{dR}
%\nonumber \\ & &
+ \left[ 1 - {\tilde {\omega }}^{2} \right] {\tilde {\omega }};
\label{win:eq:su2equations}
\end{eqnarray}
where we now have
\begin{equation}
\mu  =  1 - \frac {2{\tilde {m}}}{R} - \frac {{\tilde {\Lambda }}R^{2}}{3},
\end{equation}
and
\begin{equation}
{\tilde {G}} = \left( \frac {d{\tilde {\omega }}}{dR} \right) ^{2} , \qquad
{\tilde {p}}_{\theta } = \frac {1}{2R^{4}} \left( 1 - {\tilde {\omega }}^{2} \right) ^{2}.
\end{equation}
The equations (\ref{win:eq:su2equations}) are precisely the ${\mathfrak {su}}(2)$ EYM field equations.
Furthermore, the boundary conditions (\ref{win:eq:horizon},\ref{win:eq:infinity}) also reduce to
those for the ${\mathfrak {su}}(2)$ case.
\end{description}

\subsection{Dyonic field equations}
\label{win:sec:dyons}

As will be discussed in section \ref{win:sec:su2adSdyons}, if
either $N>2$ or we have a negative cosmological constant $\Lambda
$, then we do not need to restrict ourselves to considering only
purely magnetic equilibrium gauge potentials. If the electric part
of the gauge potential (\ref{win:eq:gaugepot}), ${\cal {A}}$, is
non-zero, there is still sufficient gauge freedom to set ${\cal
{B}}=0$ in (\ref{win:eq:gaugepot}) \cite{win:Kunzle1}. Then,
provided none of the $\omega _{j}$ vanish identically, one of the
Yang-Mills equations again tells us that all the $\gamma _{j}$ are
identically zero. Following \cite{win:Kunzle1} it is convenient to
define new real variables $\alpha _{j}(r)$ by
\begin{equation}
{\cal {A}}_{jj} = i \left[
-\frac {1}{N} \sum _{k=1}^{j-1} k \alpha _{k} + \sum _{k=j}^{N-1} \left( 1 - \frac {k}{N} \right) \alpha _{k}
\right]
\end{equation}
so that the matrix ${\cal {A}}$ is automatically purely imaginary, diagonal and traceless.
In this case the Yang-Mills equations (\ref{win:eq:YMe}) now take the form \cite{win:Kunzle1}
\begin{equation}
r^2\mu\omega''_{j}+\left(2m-2r^3 p_{\theta}-\frac{2\Lambda r^3}{3}\right)\omega'_{j}+W_j\omega_j
+\frac {\mu }{r^{2}} \alpha _{j}^{2} \omega _{j}
=0,
\end{equation}
and there are additional Yang-Mills equations for the $\alpha _{j}$, namely \cite{win:Kunzle1}
\begin{equation}
\left[ r^{2} S^{-1} \left( \mu S \alpha _{j} \right) ' \right] '
= 2\alpha _{j} \omega _{j}^{2} - \alpha _{j-1} \omega _{j-1}^{2} - \alpha _{j+1} \omega _{j+1}^{2} .
\end{equation}
The Einstein equations retain the form (\ref{win:eq:Ee}) but the
quantities $p_{\theta }$ (\ref{win:eq:ptheta}) and $G$
(\ref{win:eq:Gdef}) now read \cite{win:Kunzle1}
\begin{eqnarray}
p_{\theta } & = &
\frac{1}{4r^4}\sum^N_{j=1}\left[\left(\omega^2_j-\omega^2_{j-1}-N-1+2j\right)^2
+ \left( \frac {r^{2}}{S} \left( \mu S {\cal {A}}_{jj} \right) ' \right) ^{2}\right]
\nonumber \\
G & = & \sum^{N-1}_{j=1}\left[ \omega_j'^2 + \alpha _{j}^{2} \omega _{j}^{2} \right].
\end{eqnarray}

\subsection{Perturbation equations}
\label{win:sec:pert}

We are also interested in the stability of the static, equilibrium solutions.
For simplicity, we consider only linear, spherically symmetric perturbations of the purely magnetic solutions.
We return to the general gauge potential of the form (\ref{win:eq:gaugepot}), and the metric (\ref{win:eq:metric}),
where now all functions depend on time $t$ as well as $r$.
There is still sufficient gauge freedom to enable us to set ${\cal {A}}\equiv 0$.
This choice of gauge is particularly useful as then we shall shortly see that
the perturbation equations decouple into two sectors,
the ``gravitational'' and ``sphaleronic'' sectors \cite{win:Lavrelash1}.
We consider perturbations about the equilibrium solutions of the form
\begin{equation}
\omega _{j} (t,r) = \omega _{j} (r) + \delta \omega _{j}(t,r) ,
\end{equation}
where $\omega _{j}(r)$ are the equilibrium functions and
$\delta \omega _{j}(t,r)$ are the linear perturbations.
There are similar perturbations for the other equilibrium quantities $m$ and $S$, and in addition
we have the perturbations $\delta \gamma _{j}(t,r)$ and $\delta \beta _{j}(t,r)$, the latter being the
entries along the diagonal of the matrix ${\cal {B}}$ (\ref{win:eq:gaugepot}):
\begin{equation}
{\cal {B}} = \mbox{Diag}\left( i\delta \beta _{1} , \ldots , i\delta \beta _{N} \right) .
\end{equation}
Note that the $\delta \beta _{j}$ are not independent because the matrix ${\cal {B}}$ is traceless, so
\begin{equation}
\delta \beta _{1} + \ldots + \delta \beta _{N} =0,
\label{win:eq:deltabetaconstraint}
\end{equation}
but it simplifies the derivation of the perturbation equations to retain all the $\delta \beta _{j}$ for the moment.
We ignore all terms involving squares or higher powers of the perturbations.
The full derivation of the perturbation equations is highly involved and the details will be presented elsewhere
\cite{win:Baxter1}.
Instead here we summarize the key features of the perturbation equations.
As usual, we will employ the ``tortoise'' co-ordinate $r_{*}$, defined by:
\begin{equation}
\frac {dr_{*}}{dr} = \frac {1}{\mu S},
\label{win:eq:tortoise}
\end{equation}
where $\mu $ and $S$ are the equilibrium metric functions.

\subsubsection{Sphaleronic sector}
\label{win:sec:sphaleronic}

The sphaleronic sector consists of the $2N-1$
perturbations $\delta \beta _{j}$, $j=1,\ldots, N$ and $\delta \gamma _{j}$, $j=1, \ldots , N-1$.
We define new variables $\delta \Phi _{j}$ by
\begin{equation}
\delta \Phi _{j} = \omega _{j} \delta \gamma _{j}.
\label{win:eq:deltaPhidef}
\end{equation}
The perturbation equations for the sphaleronic sector arise solely from the Yang-Mills equations, and comprise:
\begin{eqnarray}
\delta {\ddot {\beta }}_{j} & = &
\frac {S}{r^{2}} \left[ \omega _{j-1} \partial _{r_{*}} \left( \delta \Phi _{j-1} \right)
- \omega _{j} \partial _{r_{*}} \left( \delta \Phi _{j} \right) \right]
\nonumber \\ & &
+ \frac {S}{r^{2}} \left[  \left( \partial _{r_{*}} \omega _{j} \right)  \delta \Phi _{j}
- \left( \partial _{r_{*}} \omega _{j-1} \right) \delta \Phi _{j-1} \right]
\nonumber \\ & &
+ \frac {\mu S^{2}}{r^{2}} \left[
\omega _{j}^{2} \left( \delta \beta _{j+1} - \delta \beta _{j} \right)
- \omega _{j-1}^{2} \left( \delta \beta _{j} - \delta \beta _{j-1} \right)
\right] ;
\label{win:eq:deltabeta}
\\
\delta {\ddot {\Phi }}_{j} & = &
\partial _{r_{*}}^{2} \left( \delta \Phi _{j} \right)
- \frac {1}{\omega _{j}} \left(  \partial ^{2}_{r_{*}} \omega _{j} \right) \delta \Phi _{j}
+ \mu S \omega _{j} \partial _{r_{*}} \left( \delta \beta _{j} - \delta \beta _{j+1} \right)
\nonumber \\ & &
+ \left[
\mu \left( \partial _{r_{*}} S \right) \omega _{j} + \left( \partial _{r_{*}} \mu \right) S \omega _{j}
+ 2\mu S \left( \partial _{r_{*}} \omega _{j} \right )
\right] \left( \delta \beta _{j} - \delta \beta _{j+1} \right) ;
\label{win:eq:deltaPhi}
\end{eqnarray}
together with the {\em {Gauss constraint}}
\begin{equation}
0  =  \partial _{r_{*}} \left( \delta {\dot {\beta }}_{j} \right)
+ \left[ \frac {2\mu S}{r} - \frac {\partial _{r_{*}}S}{S} \right] \delta {\dot {\beta }}_{j}
+ \frac {S}{r^{2}} \left[ \omega _{j} \delta {\dot {\Phi }}_{j}
+ \omega _{j-1} \delta {\dot {\Phi }}_{j-1} \right]  ,
\label{win:eq:GC}
\end{equation}
where a dot denotes $\partial /\partial t $.
It is important to note that the cosmological constant $\Lambda $ only appears in these equations through the
metric function $\mu $ (\ref{win:eq:mu}), and therefore the perturbation equations
(\ref{win:eq:deltabeta},\ref{win:eq:deltaPhi}) and the Gauss constraint (\ref{win:eq:GC}) have exactly the same
form as derived in \cite{win:Brodbeck1} for arbitrary gauge groups in asymptotically flat space.

\subsubsection{Gravitational sector}
\label{win:sec:gravitational}

The gravitational sector consists of the perturbations of the
metric functions $\delta \mu $ and $\delta S$ as well as the
perturbations of the remaining gauge field functions $\delta
\omega _{j}$.
Both the Einstein equations and the remaining
Yang-Mills equations are involved in this sector. For an arbitrary
gauge group and asymptotically flat space, the perturbation
equations  in this sector have been considered in
\cite{win:Brodbeck1}. In asymptotically adS, we also find that the
metric perturbations can be eliminated to give a set of equations
governing the perturbations $\delta \omega _{j}$, which can be
written in matrix form
\begin{equation}
{\underline {\delta {\ddot {\omega }}}} = \partial ^{2}_{r_{*}} \left( {\underline {\delta \omega }} \right)
+ {\cal {M}}_{G} {\underline {\delta \omega }},
\label{win:eq:deltaomega}
\end{equation}
where ${\underline {\delta \omega }} = \left(  \delta \omega _{1} , \ldots , \delta \omega _{N-1} \right) ^{T}$
and the $\left( N -1 \right) \times \left( N-1 \right) $ matrix ${\cal {M}}_{G}$ has entries
\begin{eqnarray}
{\cal {M}}_{G,j,j} & = &
\frac {\mu S^{2}}{r^{2}} \left[ W_{j} -2\omega _{j}^{2} \right]
+ \frac {4}{\mu S r} \Upsilon \left( \partial _{r_{*}} \omega _{j} \right) ^{2}
+ \frac {8S}{r^{3}} W_{j} \omega _{j} \left( \partial _{r_{*}} \omega _{j} \right) ;
\nonumber \\
{\cal {M}}_{G,j,j+1} & = &
\frac {\mu S^{2}}{r^{2}} \omega _{j} \omega _{j+1}
+\frac {4}{\mu S r} \Upsilon \left( \partial _{r_{*}} \omega _{j} \right) \left( \partial _{r_{*}} \omega _{j+1} \right)
\nonumber \\ & &
+ \frac {8S}{r^{3}} \left[ W_{j} \omega _{j} \left( \partial _{r_{*}} \omega _{j+1} \right)
+ W_{j+1} \omega _{j+1} \left( \partial _{r_{*}} \omega _{j} \right) \right] ;
\nonumber \\
{\cal {M}}_{G,j,k} & = &
\frac {4}{\mu S r} \Upsilon \left( \partial _{r_{*}} \omega _{j} \right) \left( \partial _{r_{*}} \omega _{k} \right)
+ \frac {8S}{r^{3}} \left[ W_{j} \omega _{j} \left( \partial _{r_{*}} \omega _{k} \right)
+ W_{k} \omega _{k} \left( \partial _{r_{*}} \omega _{j} \right) \right] ;
\label{win:eq:calMG}
\end{eqnarray}
where $k \neq j, j+1$, and $\Upsilon $ is given in terms of the equilibrium metric functions $\mu $ and $S$ as follows:
\begin{equation}
\Upsilon = \frac {1}{\mu } \partial _{r_{*}} \mu + \frac {1}{S} \partial _{r_{*}} S + \frac {\mu S}{r}.
\end{equation}

\section{Asymptotically flat/de Sitter solutions for ${\mathfrak {su}}(N)$ EYM}
\label{win:sec:af}

We now turn to black hole solutions of the EYM field equations, beginning by briefly reviewing some of the
key features of solutions in asymptotically flat or asymptotically de Sitter space.

\subsection{Asymptotically flat, spherically symmetric ${\mathfrak {su}}(2)$ solutions}
\label{win:sec:afsu2}

Apart from the trivial solutions given above (\ref{win:eq:SadS},\ref{win:eq:RNadS}),
the first black hole solutions of the EYM field equations were found by Yasskin \cite{win:Yasskin},
and correspond to embedding the Reissner-Nordstr\"om electromagnetic gauge field into a higher-dimensional
gauge group.
The metric of these solutions is still Reissner-Nordstr\"om.
Yasskin conjectured that his solutions were the only ones possible.
This conjecture was only shown to be false twenty-five years later \cite{win:Bizon1,win:Kunzle3,win:Volkov2,win:Volkov3}.
That the discovery of hairy black holes in ${\mathfrak {su}}(2)$ EYM took so long may be attributed
to the conjecture that there were no soliton solutions in this model.
This conjecture is based on the fact that there are no solitons in pure gravity (see, for example,
\cite{win:Heusler3,win:Lichnerowicz1});
no solitons in Einstein-Maxwell theory \cite{win:Heusler1}, no pure YM solitons in flat space-time
\cite{win:Coleman1,win:Deser1} and
no EYM solitons in three space-time dimensions \cite{win:Deser2}.
However, once Bartnik and McKinnon \cite{win:Bartnik2} had discovered non-trivial EYM solitons in four-dimensional
space-time, Yasskin's no-hair conjecture for EYM theory was quickly shown to be false \cite{win:Bizon1}.

For ${\mathfrak {su}}(2)$ EYM, it has been shown \cite{win:Bizon5,win:Ershov2,win:Ershov1}  that non-trivial solutions
(i.e. solutions in which the gauge field is not essentially Abelian) must have a purely magnetic gauge potential,
which is described by a single gauge field function $\omega (r)$ (\ref{win:eq:gaugepotsimple}).
Note that the ansatz (\ref{win:eq:gaugepotsimple}) for ${\mathfrak {su}}(2)$
is not the same as the Witten ansatz \cite{win:Witten1} which was used in the original papers
\cite{win:Bartnik2,win:Bizon1}, but it gives equivalent field equations.
In this case the ${\mathfrak {su}}(2)$ EYM equations have the form
\begin{eqnarray}
\frac {dm}{dr} & = &
 \left( 1 -\frac {2m}{r} \right)  \left( \frac {d\omega }{dr} \right) ^{2}
+ \frac {1}{2r^{2}} \left( 1- \omega ^{2} \right) ^{2} ; \qquad
\nonumber \\
\frac {1}{S} \frac {dS}{dr} & = &
- \frac {2}{r} \left( \frac {d\omega }{dr} \right) ^{2} ;
\nonumber \\
0 & = & r^{2} \left( 1- \frac {2m}{r} \right)  \frac {d^{2}\omega }{dr^{2}} +
\left[ 2m - \frac {\left( 1- \omega ^{2} \right) ^{2}}{r} \right] \frac {d\omega }{dr}
%\nonumber \\ & &
+ \left[ 1 - \omega ^{2} \right] \omega .
\label{win:eq:su2equationsAF}
\end{eqnarray}
It is the highly non-linear nature of these equations which allows for non-trivial soliton and hairy black hole
solutions, which may be thought of heuristically as arising from a balancing of the gravitational and gauge field
interactions (see \cite{win:Hod1} for a recent discussion).
The non-linear nature of the equations also means, however, that (apart from the solutions for the Yang-Mills
field on a fixed Schwarzschild metric \cite{win:Boutaleb1,win:Brihaye8}) solutions can only be found numerically.

The numerical work in \cite{win:Bartnik2,win:Bizon1,win:Kunzle3,win:Volkov2,win:Volkov3}
found discrete families of solutions \cite{win:Sudarsky2}, indexed by the event horizon radius $r_{h}$ (with $r_{h}=0$ for
solitons) and $n$, the number of zeros of the single gauge field function $\omega $, each pair
$\left( r_{h},n \right)$ identifying a solution of the field equations.
A key feature of the solutions is that $n>0$, so that the gauge field function must have at least one zero
(or ``node'').
Later analytic work \cite{win:Breitenlohner3,win:Smoller2,win:Smoller3,win:Smoller1} rigorously proved
these numerical features.
The black holes are ``hairy'' in the sense that they have no
magnetic charge \cite{win:Bizon5,win:Ershov2,win:Ershov1}, and are therefore indistinguishable at infinity
from a standard Schwarzschild black hole.
However, the ``hair'', that is, the non-trivial structure in the matter fields, extends some way out
from the event horizon, leading to the ``no-short-hair'' conjecture \cite{win:Nunez1}.

Although initially controversial \cite{win:Bizon3,win:Bizon4,win:Straumann4,win:Wald1},
rapidly it was accepted that both the soliton \cite{win:Straumann1} and black hole solutions \cite{win:Straumann2}
are unstable.
This instability is not unexpected if we consider the solutions as arising from a balancing of the gauge field
and gravitational interactions.
Studies of the non-linear stability of the solutions \cite{win:Zhou1,win:Zhou2} reveal that the gauge field ``hair''
either radiates away to infinity or falls down the black hole event horizon, leaving, as the end-point, a bald
Schwarzschild black hole.
Due to this instability, the black holes,
while they violate the ``letter'' of the no-hair conjecture, may be thought of
as not contradicting its ``spirit'', and one might be led to conjecture that all {\em {stable}} black holes
are fixed by their mass, angular momentum, and conserved charges.

Originally these hairy black holes were shown to be unstable using numerical techniques \cite{win:Straumann2}
but the instability can also be shown analytically \cite{win:Galtsov3,win:Volkov4}.
In the ${\mathfrak {su}}(2)$ case, the perturbation equations
(\ref{win:eq:deltabeta},\ref{win:eq:deltaPhi},\ref{win:eq:deltaomega}) simplify considerably.
The sphaleronic sector reduces to a single equation (see section \ref{win:sec:su2adSstab} below for further details)
\begin{equation}
-{\ddot {\zeta }} = -\partial _{r_{*}}^{2} \zeta  + \left[
\frac {\mu S^{2}}{r^{2}} \left( 1 + \omega ^{2} \right) + \frac {2}{\omega ^{2}} \left(
\frac {d\omega }{dr_{*}} \right) ^{2}
\right] \zeta ,
\label{win:eq:sphalsu2AF}
\end{equation}
while, on eliminating the metric perturbations, the gravitational sector also has just one equation:
\begin{eqnarray}
-\delta {\ddot {\omega }} &  = &
-\partial _{r_{*}}^{2} \left( \delta \omega \right)
\label{win:eq:gravsu2AF}
\\ & &
+ \frac {\mu S^{2}}{r^{2}} \left[
3\omega ^{2} - 1 -4r\omega '^{2} \left( \frac {1}{r}
- \frac {\left( 1- \omega ^{2} \right) ^{2}}{r^{3}} \right)
+ \frac {8}{r} \omega \omega ' \left( \omega ^{2}-1 \right)
\right] \delta \omega .
\nonumber
\end{eqnarray}
The instability has been compared to that of the flat-space Yang-Mills sphaleron \cite{win:Volkov4},
which has a single unstable mode.
The situation is slightly more complicated here, due to the two sectors of perturbations.
The sphaleronic sector certainly, as its name suggests, mimics the perturbations of the flat-space sphaleron.
It can be shown \cite{win:Volkov5} that the number of instabilities in the sphaleronic sector equals
$n$, the number of zeros of the gauge field function $\omega $.
The same is true in the gravitational sector, as conjectured in \cite{win:Lavrelash1} and can be
shown using catastrophe theory, by considering the more general EYM-Higgs solutions \cite{win:Mavromatos3}.
The above concerns only spherically symmetric perturbations.
It is known that the flat-space sphaleron has instabilities only in the spherically symmetric sector \cite{win:Baacke1}.
Extending this to the ${\mathfrak {su}}(2)$ EYM black holes requires complicated analysis \cite{win:Sarbach4},
using a curvature-based formalism developed in \cite{win:Brodbeck2,win:Sarbach3,win:Sarbach2}.

Using the isolated horizons formalism,  these ``hairy'' black holes can be interpreted as bound states
of ordinary black holes with the Bartnik-MacKinnon solitons \cite{win:Ashtekar1,win:Corichi2,win:Corichi1}.
In particular, the soliton masses are given in terms of the masses of the corresponding black holes \cite{win:Corichi1},
and the instability of the coloured black holes arises naturally from the instability of the corresponding solitons
\cite{win:Ashtekar1,win:Corichi2}.

Since these initial discoveries
a plethora of new, asymptotically flat,
hairy black hole solutions have been found in Einstein-Yang-Mills theory and its variants
(see \cite{win:Volkov1} for a review
of those solutions discovered prior to 1999).
Most of these are, indeed, unstable.
However, there are notable exceptions, including: (a) the Skyrme black hole
\cite{win:Bizon2,win:Droz1,win:Droz2,win:Heusler4}
where the existence of an integer-valued topological winding number renders the solutions stable,
(b) Einstein-Yang-Mills-Higgs black holes in the limit of infinitely strong coupling of the Higgs field
\cite{win:Aichelburg1}
and (c) a particular branch of Einstein-non-Abelian-Proca black holes
\cite{win:Greene1,win:Maeda1,win:Torii5,win:Tamaki1,win:Torii4}.
We will not consider additional matter fields further in this article.

\subsection{Non-spherically symmetric, asymptotically flat ${\mathfrak {su}}(2)$ solutions}
\label{win:sec:nonss}

One of the surprising aspects of the failure of black hole uniqueness in EYM is that
almost every step in the uniqueness theorem in Einstein-Maxwell theory has a counter-example in EYM
(see \cite{win:Heusler2} for detailed discussions on this topic, and
\cite{win:Brodbeck6,win:Racz1,win:Straumann3,win:Sudarsky2,win:Sudarsky3}
for examples of some results from Einstein-Maxwell theory which do generalize).
An important example of this is Israel's theorem \cite{win:Israel1,win:Israel2},
which states that the geometry outside the event horizon of a static black hole must be spherically symmetric.
This is not true in EYM: there are static black hole solutions which are not spherically symmetric
but only axisymmetric \cite{win:Kleihaus4}
(in more general matter models, static black holes do not necessarily possess any symmetries at all
\cite{win:Ridgway1,win:Ridgway2}).
These solutions are found numerically by writing the metric in isotropic co-ordinates:
\begin{equation}
ds^{2} = -f(r,\theta ) \, dt^{2} + \frac {m(r,\theta )}{f(r,\theta )} dr^{2}
+ \frac {m(r,\theta ) r^{2}}{f(r,\theta )} d\theta ^{2}
+ \frac {L(r,\theta ) r^{2} \sin ^{2} \theta }{f(r,\theta )} d\phi ^{2},
\label{win:eq:metriciso}
\end{equation}
and using the following ansatz for the ${\mathfrak {su}}(2)$ gauge field \cite{win:Rebbi1}:
\begin{eqnarray}
A & = & \frac {1}{2r} \left\{
\tau _{\phi }^{p} \left[ H_{1}(r,\theta ) \, dr + \left( 1 - H_{2}(r,\theta ) \right) r \, d\theta \right]
\right. \nonumber \\ & &  \left.
- p \left[ \tau _{r}^{p} H_{3} (r,\theta ) + \tau _{\theta }^{p} \left( 1- H_{4} (r,\theta ) \right) \right]
r \sin \theta \, d\phi
\right\} ,
\label{win:eq:gaugeiso}
\end{eqnarray}
where
\begin{eqnarray}
\tau _{r}^{p} & = &
 {\underline {\tau .}} \left( \sin \theta \cos p\phi , \sin \theta \sin p\phi , \cos \theta \right) ,
\nonumber \\
\tau _{\theta }^{p} & = &
{\underline {\tau .}} \left( \cos \theta \cos p\phi , \cos \theta \sin p\phi , - \sin \theta \right) ,
\nonumber \\
\tau _{\phi }^{p} & = &
{\underline {\tau .}} \left( -\sin p\phi , \cos p\phi , 0 \right) ,
\label{win:eq:winding}
\end{eqnarray}
with
\begin{equation}
{\underline {\tau }} = \left( \tau _{x}, \tau _{y}, \tau _{z} \right) ,
\end{equation}
where $\tau _{x}$, $\tau _{y}$, $\tau _{z}$ are the usual generators of ${\mathfrak {su}}(2)$.
Here, $p$ is a winding number, with $p=1$ corresponding to spherically symmetric solutions (with the gauge
potential written in a different form to that we have used in (\ref{win:eq:gaugepotsimple})).
Substituting the ansatz into the field equations gives a complicated set of partial differential equations,
solutions of which are exhibited in \cite{win:Kleihaus4}.
Static, axisymmetric soliton solutions also exist \cite{win:Galtsov4,win:Ibadov1,win:Kleihaus5}.

It is less surprising that rotating black holes also exist in this model \cite{win:Kleihaus6,win:Kleihaus7},
generalizing the Kerr-Newman metric (as predicted in \cite{win:Sudarsky2}).
These solutions are indexed by the winding number $p$ (\ref{win:eq:winding}) and a node number $n$.
They carry no magnetic charge, but all have non-zero electric charge \cite{win:Sudarsky2,win:Sudarsky3}.
The question of whether there are rotating solitons in pure ${\mathfrak {su}}(2)$ EYM has yet to be
conclusively settled, however.
Rotating soliton solutions have been found in EYM-Higgs theory \cite{win:Paturyan1},
but not in pure EYM theory.
Although rotating solitons are predicted perturbatively \cite{win:Brodbeck4},
the consensus in the literature is now that it seems unlikely that
rotating soliton solutions do exist \cite{win:Bij3}.

\subsection{Asymptotically flat ${\mathfrak {su}}(N)$ solutions}
\label{win:sec:afsuN}

We shall next consider generalizations of the ${\mathfrak {su}}(2)$ YM gauge group.
The simplest such generalization is to consider ${\mathfrak {su}}(N)$ EYM.
The results of \cite{win:Ershov2,win:Ershov1} do not extend to this larger gauge group, and it
is possible to have solutions with electric charge \cite{win:Galtsov1},
which correspond to a superposition of electrically charged Reissner-Nordstr\"om and the
${\mathfrak {su}}(2)$ EYM black holes.
Numerical solutions of the field equations have been found in the following papers:
\cite{win:Galtsov1,win:Kleihaus1,win:Kleihaus2,win:Kleihaus3}.
As $N$ increases, the possible structures of the gauge field potential (\ref{win:eq:gaugepot})
become ever more complicated.
A method for computing all spherically symmetric ${\mathfrak {su}}(N)$
gauge field potentials is given in \cite{win:Bartnik1}, where all the irreducible possibilities are enumerated
for $N\le 6$.
As in the ${\mathfrak {su}}(2)$ case, black hole solutions are found at discrete points in the parameter space
$ \{ \omega _{j}(r_{h}),j=1\ldots N-1 \} $.

There is comparatively little analytic work for more general gauge groups.
Local existence of solutions of the field equations (\ref{win:eq:YMe},\ref{win:eq:Ee}) near the black hole
event horizon and at infinity has been proven for gauge group ${\mathfrak {su}}(N)$ \cite{win:Kunzle2},
and subsequently extended to arbitrary compact gauge group \cite{win:Oliynyk1,win:Oliynyk2}.
The existence of non-trivial black hole solutions to the field equations has been proven rigorously
only in the ${\mathfrak {su}}(3)$ case \cite{win:Ruan1,win:Ruan2}, although there are arguments
that hairy black hole solutions exist for all $N$ \cite{win:Mavromatos2}.
In the ${\mathfrak {su}}(3)$ case, Ruan \cite{win:Ruan1,win:Ruan2} has proved that there are infinitely many
hairy black hole solutions, indexed by the numbers of zeros $(n_{1},n_{2})$ respectively of the two gauge
field functions $(\omega _{1},\omega _{2})$.
Furthermore, provided that the radius of the event horizon is sufficiently large, there is a black hole
solution for any combination of $(n_{1},n_{2})$.
The global properties of the solutions for arbitrary compact gauge group are studied in \cite{win:Oliynyk3}.
However, it will come as no surprise to learn that all these solutions, in asymptotically flat space, and for
any compact gauge group, are unstable \cite{win:Brodbeck3,win:Brodbeck1}.
To show instability it is sufficient to find a single unstable mode, and therefore the work in
\cite{win:Brodbeck3,win:Brodbeck1} studies the simpler, sphaleronic sector of perturbations
(see section \ref{win:sec:sphaleronic}).

\subsection{Asymptotically de Sitter ${\mathfrak {su}}(2)$ EYM solutions}
\label{win:sec:dS}

Another natural generalization of asymptotically flat ${\mathfrak {su}}(2)$ EYM
is the inclusion of a non-zero cosmological constant $\Lambda $.
When the cosmological constant is positive, soliton \cite{win:Volkov6} and black hole \cite{win:Torii3}
${\mathfrak {su}}(2)$ EYM solutions have been found (other numerical solutions are presented in
\cite{win:Brihaye9,win:Molnar1}).
These solutions possess a cosmological horizon and approach de Sitter space at infinity
(for a complete classification of the possible space-time structures, see \cite{win:Breitenlohner1}).
The phase space of solutions is again discrete, and the single gauge field function
$\omega $ must have at least one zero.
Unsurprisingly, these solutions again turn out to be unstable \cite{win:Brodbeck5,win:Forgacs1,win:Torii3}.
Given this instability, the asymptotically de Sitter solutions have received rather less attention in the
literature, but some analytic work can be found in \cite{win:Linden3,win:Linden2,win:Linden1}.

\section{Asymptotically anti-de Sitter solutions for ${\mathfrak {su}}(2)$ EYM}
\label{win:sec:su2adS}

We now turn to the main focus of this article: asymptotically anti-de Sitter solutions.
We begin by reviewing some of the properties of black holes in ${\mathfrak {su}}(2)$ EYM.

\subsection{Spherically symmetric, asymptotically adS, ${\mathfrak {su}}(2)$ EYM solutions}
\label{win:sec:su2adSnum}

Black hole solutions of ${\mathfrak {su}}(2)$ EYM with a negative cosmological constant
were first studied in \cite{win:ew1}, and subsequently in \cite{win:Bjoraker1,win:Bjoraker2}.
The field equations now take the form
\begin{eqnarray}
\frac {dm}{dr} & = &
 \left( 1 -\frac {2m}{r} -\frac {\Lambda r^{2}}{3} \right)  \left( \frac {d\omega }{dr} \right) ^{2}
+ \frac {1}{2r^{2}} \left( 1- \omega ^{2} \right) ^{2} ; \qquad
\nonumber \\
\frac {1}{S} \frac {dS}{dr} & = &
- \frac {2}{r} \left( \frac {d\omega }{dr} \right) ^{2} ;
\nonumber \\
0 & = & r^{2} \left( 1- \frac {2m}{r} -\frac {\Lambda r^{2}}{3} \right)  \frac {d^{2}\omega }{dr^{2}} +
\left[ 2m - \frac {2\Lambda r^{3}}{3} - \frac {\left( 1- \omega ^{2} \right) ^{2}}{r} \right] \frac {d\omega }{dr}
\nonumber \\ & &
+ \left[ 1 - \omega ^{2} \right] \omega .
\label{win:eq:su2equationsAdS}
\end{eqnarray}
The inclusion of a negative cosmological constant means that boundary conditions at infinity
(\ref{win:eq:infinity}) are considerably less stringent than in the asymptotically flat case;
it is therefore unsurprising that
it is easier to find solutions in asymptotically adS.

\begin{figure}
\includegraphics[width=8cm]{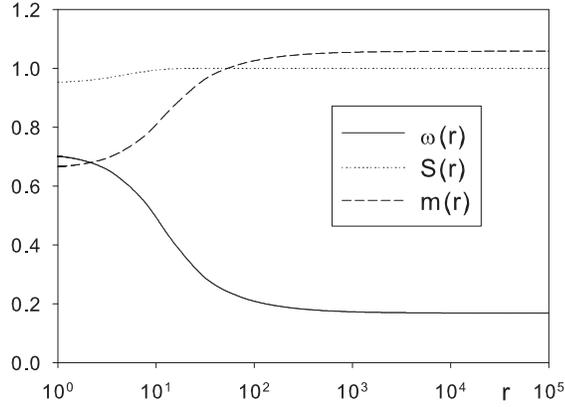}
\caption{An example of an ${\mathfrak {su}}(2)$ EYM black hole in adS in which the gauge field function $\omega (r)$
has no zeros.
Here, $\Lambda = -1$, $r_{h}=1$ and $\omega (r_{h})=0.7$.}
\label{win:fig:su2example}
\end{figure}

The space of solutions in adS is very different to that in asymptotically flat space.
Instead of finding solutions at discrete values of $\omega (r_{h})$, instead solutions exist in continuous, open
intervals.
Furthermore, for sufficiently large $\left| \Lambda \right| $,
we now find solutions in which the single gauge field function $\omega (r)$ has no zeros.
A typical example of such a solution is shown in figure \ref{win:fig:su2example}, further
examples can be found in \cite{win:ew1}.
These properties of the space of solutions of the equations (\ref{win:eq:su2equationsAdS}) are proved in \cite{win:ew1}.

\begin{figure}
\includegraphics[angle=270,width=8cm]{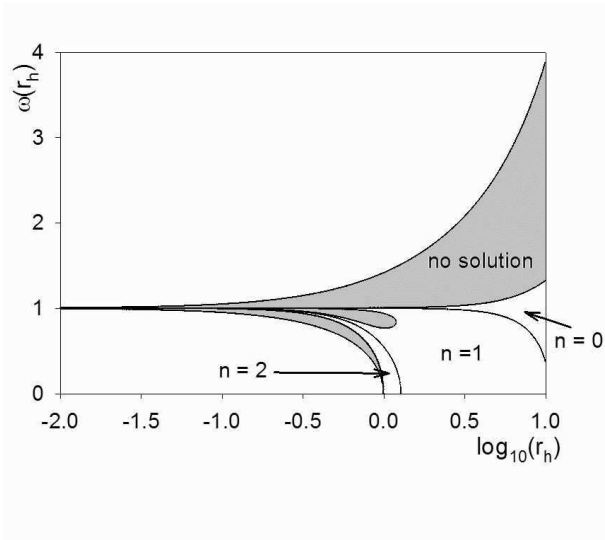}
\caption{The space of ${\mathfrak {su}}(2)$ black hole solutions when $\Lambda = -0.01$, for varying $r_{h}$.
The shaded region indicates values of the gauge field function $\omega (r_{h})$ at the event horizon
for which the constraint (\ref{win:eq:su2bhconstraint}) is satisfied, but for which we find no well-behaved black
hole solution.  The number of zeros $n$ of the gauge field function $\omega $ are indicated in those regions
of the phase space where we find black hole solutions. Elsewhere on the diagram, the constraint
(\ref{win:eq:su2bhconstraint}) is not satisfied.
Between the region where $n=2$ and the shaded region we find black hole solutions with $n=3$, $4$ and $5$,
but these regions are too small to indicate on the graph.
Taken from \cite{win:Baxter3}.}
\label{win:fig:su2bh1}
\end{figure}

We now examine the structure of the space of solutions, more details of which can be found in
\cite{win:Baxter2,win:Baxter3,win:ew1}.
There are three parameters describing the solutions, $r_{h}$, $\Lambda $ and $\omega (r_{h})$.
In order to plot two-dimensional figures, we fix either $r_{h}$ or $\Lambda $ and vary the other two quantities.
For ${\mathfrak {su}}(2)$ black holes, the constraint (\ref{win:eq:constraint}) on the value of the gauge field
function at the event horizon reads
\begin{equation}
\left( \omega (r_{h}) ^{2} - 1 \right) ^{2} < r_{h}^{2} \left( 1 - \Lambda r_{h}^{2} \right) .
\label{win:eq:su2bhconstraint}
\end{equation}
Whether we are varying $r_{h}$ or $\Lambda $, we perform a scan over all values of $\omega _{h}$ which
satisfy (\ref{win:eq:su2bhconstraint}).
Firstly, we show in figure \ref{win:fig:su2bh1} the space of black hole solutions for fixed $\Lambda = -0.01$ and
varying event horizon radius $r_{h}$.
The outermost curves in figure \ref{win:fig:su2bh1} are where the inequality (\ref{win:eq:su2bhconstraint})
is saturated.
Immediately inside these curves we have a shaded region,
which represents values of $\left( r_{h}, \omega (r_{h}) \right) $
for which the constraint (\ref{win:eq:su2bhconstraint}) is satisfied, but for which we are unable to find
black hole solutions which remain regular all the way out to infinity.
Where we do find solutions, we indicate in figure~\ref{win:fig:su2bh1} the number of
zeros of the gauge field function $\omega (r)$.
The solution for which $\omega (r_{h})=1$ is simply the Schwarzschild-adS black hole, while that for $\omega (r_{h})=0$
is the magnetically charged Reissner-Nordstr\"om-adS black hole (see section \ref{win:sec:trivial}).
As $r_{h}\rightarrow 0$, the constraint (\ref{win:eq:su2bhconstraint}) implies that $\omega (r_{h}) \rightarrow 1$,
as can be seen in figure \ref{win:fig:su2bh1}.
The black hole solutions become solitons in this limit.
However, for this value of $\Lambda$, there are different
soliton solutions, with $\omega $ having different numbers of zeros \cite{win:Breitenlohner2}, a
feature which is not readily apparent from figure \ref{win:fig:su2bh1}.
We find similar behaviour on varying $r_{h}$ for different values of $\Lambda $.

\begin{figure}
\includegraphics[angle=270,width=9cm]{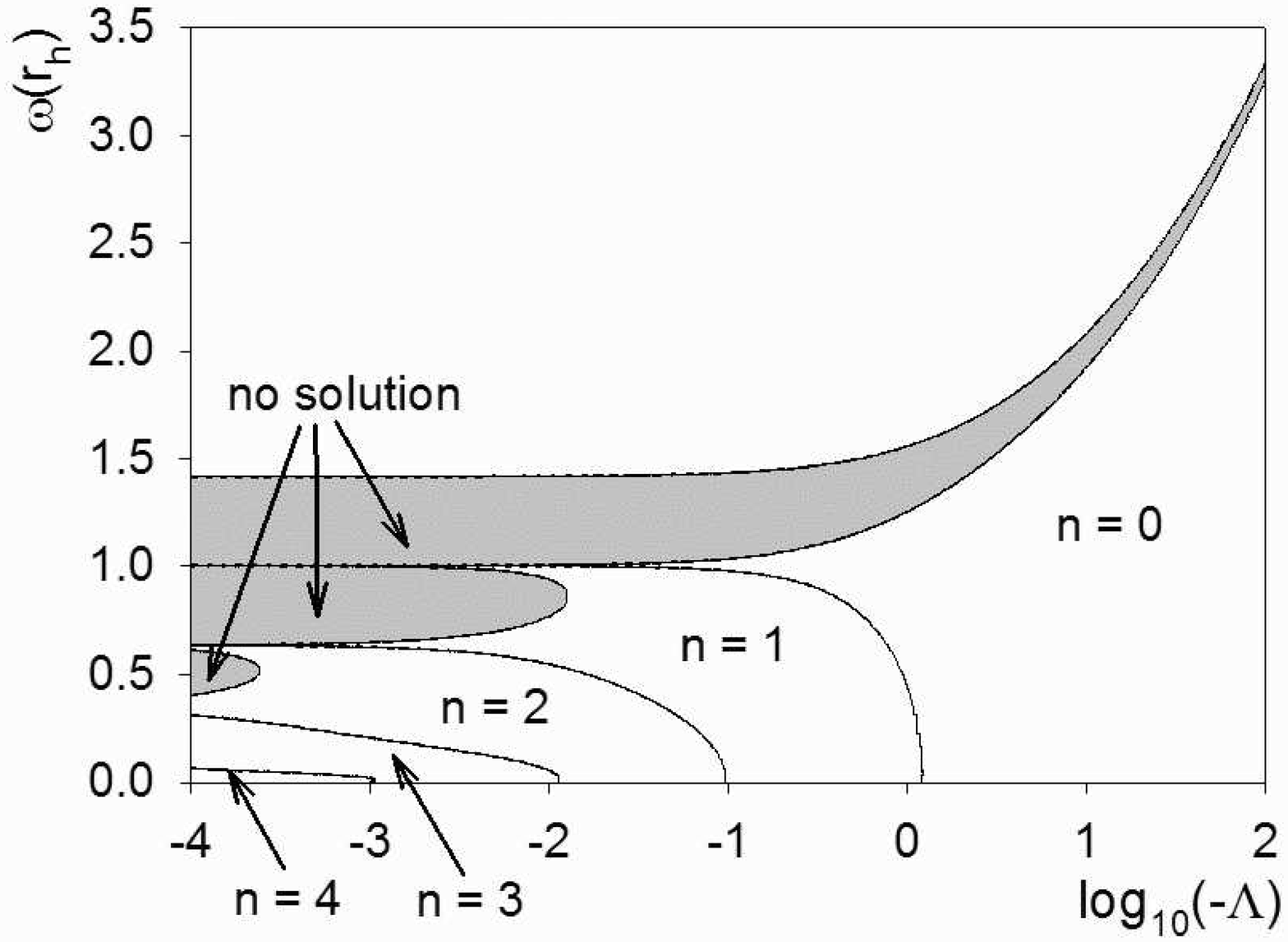}
\caption{Phase space of ${\mathfrak {su}}(2)$ black holes with $r_{h}=1$ and varying $\Lambda $.
The shaded region indicates values of the gauge field function $\omega (r_{h})$ at the event horizon
for which the constraint (\ref{win:eq:su2bhconstraint}) is satisfied, but for which we find no well-behaved black
hole solution.  The number of zeros $n$ of the gauge field function $\omega $ are indicated in those regions
of the phase space where we find black hole solutions. Elsewhere on the diagram, the constraint
(\ref{win:eq:su2bhconstraint}) is not satisfied.
As well as the regions where $n=0,\ldots,4$ as marked on the diagram, we find a small region in
the bottom left of the plot where $n=5$. This region is too small to indicate on the current figure,
but can be seen in figure \ref{win:fig:su2bh3}.
Taken from \cite{win:Baxter3}.}
\label{win:fig:su2bh2}
\end{figure}

\begin{figure}
\includegraphics[angle=270,width=9cm]{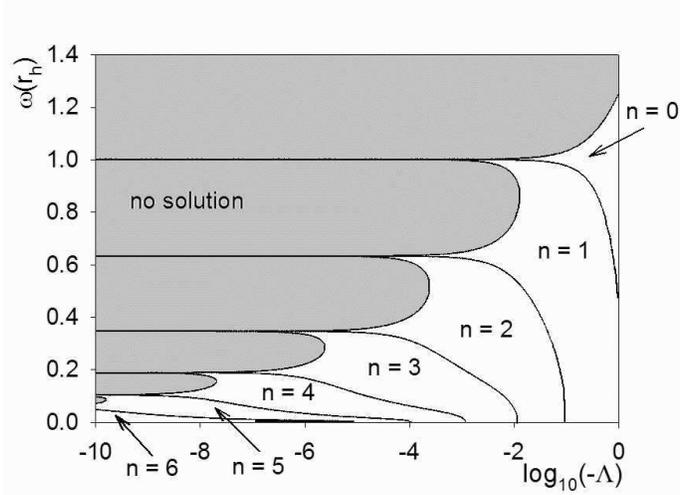}
\caption{Close-up of the phase space of ${\mathfrak {su}}(2)$ black holes with $r_{h}=1$ and smaller values
of $\Lambda $.
In the bottom left of the plot there is a small region of solutions for which $n=7$, but the region
is too small to be visible.
Taken from \cite{win:Baxter3}.}
\label{win:fig:su2bh3}
\end{figure}

If we now fix the event horizon radius to be $r_{h}=1$ and vary $\Lambda $,
the solution space is shown in figure \ref{win:fig:su2bh2}, with a close-up for smaller values of
$\left| \Lambda \right| $ in figure \ref{win:fig:su2bh3}.
Again, in figures \ref{win:fig:su2bh2} and \ref{win:fig:su2bh3} we have shaded those regions where the constraint
(\ref{win:eq:su2bhconstraint}) is satisfied, but no regular black hole solutions could be found.
Where we do find solutions, the number of zeros of the gauge field function $\omega (r)$ is indicated in the figures.
As $\Lambda \rightarrow 0$, the phase space breaks up into discrete points, which correspond to the asymptotically
flat `colored' ${\mathfrak {su}}(2)$ black holes described in section \ref{win:sec:afsu2} \cite{win:Bizon1}.
For sufficiently large $\left| \Lambda \right| $, we find solutions in which the gauge field function has no zeros.

\begin{figure}
\includegraphics[angle=270,width=11cm]{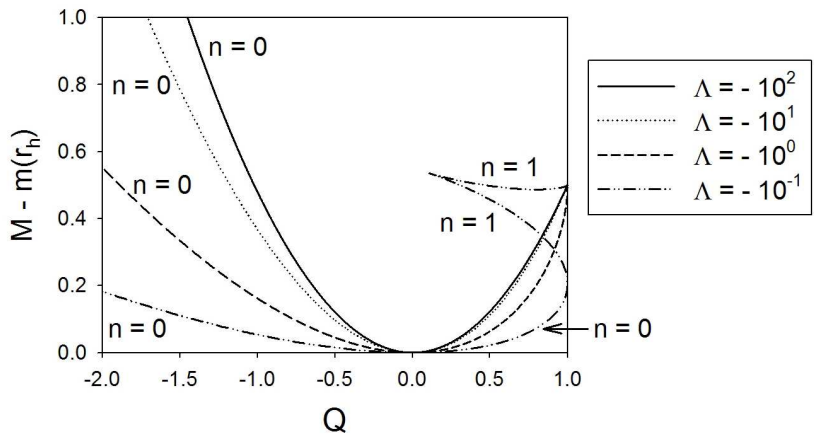}
\caption{Black hole mass $M$ and magnetic charge $Q$ for ${\mathfrak {su}}(2)$ EYM black holes
with $r_{h}=1$ and varying $\Lambda $ (cf. figure 8 in \cite{win:Bjoraker2}).}
\label{win:fig:su2charge1}
\end{figure}

The spectrum of black hole solutions (that is, the relationship between the mass $M$ and
magnetic charge $Q$ of the black holes) was first studied in \cite{win:Bjoraker2}.
We plot in figure \ref{win:fig:su2charge1} the black hole mass versus magnetic charge for
black holes with $r_{h}=1$ and varying values of $\Lambda $ (cf. figure 8 in \cite{win:Bjoraker2}).
For large values of $\left| \Lambda \right| $, there are only nodeless solutions and the spectrum
is simple, with the black holes being uniquely specified by $\Lambda $, $r_{h}$ and $Q_{M}$.
As $\left| \Lambda \right| $ decreases, the spectrum becomes more complicated.
For example, looking at the $\Lambda = -0.1$ curve in figure \ref{win:fig:su2charge1} we see that a branch
structure emerges.
The lower $M$ curve for $\Lambda = -0.1$ consists of $n=0$ (nodeless) solutions, and extends from negative $Q$
up to $Q=1$.
When $Q=1$, a branch of $n=1$ solutions appears, which have larger mass.
As $Q$ decreases along this branch of solutions, the mass $M$ increases, until a bifurcation point is
reached and a second branch of $n=1$ solutions appears, with even larger mass, and with the charge increasing as $M$
increases.
For smaller values of $\left| \Lambda \right| $, we find ever more complicated spectra, which appear
to become ``fractal'' as $\left| \Lambda \right| \rightarrow 0$ \cite{win:Bjoraker2,win:Matinyan1}.
In view of the catastrophe theory analysis of other hairy black hole solutions \cite{win:Torii5,win:Tamaki1,win:Torii4},
one might anticipate that the stability of the solutions changes at the points in the spectrum where two branches
of solutions meet, but this has yet to be fully investigated in the literature
(see \cite{win:Breitenlohner2} for an in-depth stability analysis of the soliton solutions).
We therefore next consider the stability of these black holes.

\subsection{Stability of the spherically symmetric solutions}
\label{win:sec:su2adSstab}

As discussed in section \ref{win:sec:afsu2}, for the asymptotically flat ${\mathfrak {su}}(2)$ EYM black holes,
it has been shown that the number of instabilities is twice the number of zeros of the gauge field function $\omega (r)$.
Therefore, one might anticipate that at least some solutions when $\omega (r)$ has no zeros could be stable.
For the ${\mathfrak {su}}(2)$ EYM case, the perturbation equations
(\ref{win:eq:deltabeta},\ref{win:eq:deltaPhi},\ref{win:eq:deltaomega}) simplify considerably.
In the sphaleronic sector, there is a single $\delta \Phi $ (\ref{win:eq:deltaPhidef}) and two further perturbations
$\delta \beta _{1}$, $\delta \beta _{2}$, although these are not independent (\ref{win:eq:deltabetaconstraint}),
so we may consider just $\delta \nu  = \delta \beta _{2} - \delta \beta _{1}$.
The sphaleronic sector perturbations equations (\ref{win:eq:deltabeta},\ref{win:eq:deltaPhi}) then reduce to
\begin{eqnarray}
\delta {\ddot {\nu }} & = &
\frac {2S}{r^{2}} \left[ \omega  \partial _{r_{*}} \left( \delta \Phi  \right)
- \left( \partial _{r_{*}} \omega  \right) \delta \Phi \right]
- \frac {2\mu S^{2}}{r^{2}} \omega ^{2} \delta \nu ;
\\
\delta {\ddot {\Phi }} & = &
\partial _{r_{*}}^{2} \left( \delta \Phi \right)
- \frac {1}{\omega } \left(  \partial ^{2}_{r_{*}} \omega  \right) \delta \Phi
- \mu S \omega \partial _{r_{*}} \left( \delta \nu \right)
\nonumber \\ & &
+ \left[
\mu \left( \partial _{r_{*}} S \right) \omega  + \left( \partial _{r_{*}} \mu \right) S \omega
- 2\mu S \left( \partial _{r_{*}} \omega  \right )
\right] \delta \nu  ;
\end{eqnarray}
and the Gauss constraint (\ref{win:eq:GC}) is now
\begin{equation}
0  =  \partial _{r_{*}} \left( \delta {\dot {\nu }} \right)
+ \left[ \frac {2\mu S}{r} - \frac {\partial _{r_{*}}S}{S} \right] \delta {\dot {\nu }}
+ \frac {S}{r^{2}} \omega \delta {\dot {\Phi }} .
\end{equation}
By introducing a new variable $\zeta $ by (note our notation above is different from that used in \cite{win:ew1})
\begin{equation}
\zeta = \frac {r^{2}}{S} \delta \nu ,
\end{equation}
the sphaleronic sector then reduces to a single equation \cite{win:ew1}
\begin{equation}
-{\ddot {\zeta }} = -\partial _{r_{*}}^{2}\zeta + \left[
\frac {\mu S^{2}}{r^{2}} \left( 1 + \omega ^{2} \right) + \frac {2}{\omega ^{2}} \left(
\frac {d\omega }{dr_{*}} \right) ^{2}
\right] \zeta ,
\label{win:eq:sphalsu2AdS}
\end{equation}
while the gravitational sector (\ref{win:eq:deltaomega}) also has just one equation:
\begin{eqnarray}
-\delta {\ddot {\omega }} & =  &
-\partial _{r_{*}}^{2} \left( \delta \omega \right)
\label{win:eq:gravsu2AdS}
\\ & &
+ \frac {\mu S^{2}}{r^{2}} \left[
3\omega ^{2} - 1 -4r\omega '^{2} \left( \frac {1}{r} - \Lambda r
- \frac {\left( 1- \omega ^{2} \right) ^{2}}{r^{3}} \right)
+ \frac {8}{r} \omega \omega ' \left( \omega ^{2}-1 \right)
\right] \delta \omega .
\nonumber
\end{eqnarray}
The sphaleronic sector equation (\ref{win:eq:sphalsu2AdS}) is exactly the same as that in the asymptotically flat
${\mathfrak {su}}(2)$ EYM case (\ref{win:eq:sphalsu2AF}), but the gravitational sector equation (\ref{win:eq:gravsu2AF})
unsurprisingly is modified by the presence of non-zero $\Lambda $.
Both equations (\ref{win:eq:sphalsu2AdS}) and (\ref{win:eq:gravsu2AdS}) have the standard Schr\"odinger form
\begin{equation}
-{\ddot {\Psi }} = -\partial _{r_{*}}^{2}  \Psi  + {\cal {U}} \Psi ,
\end{equation}
with potential ${\cal {U}}$.
For the sphaleronic sector, when the gauge field function $\omega (r)$ has no zeros, it is immediately clear
that the potential ${\cal {U}}$ is positive, so there are no instabilities in this sector
(this result does not hold in the asymptotically flat case because the zeros of $\omega (r)$ in that
case mean that ${\cal {U}}$ is not regular).
The gravitational sector potential is more complex to analyze, but, for sufficiently large $\left| \Lambda \right| $
and $\omega (r_{h})> 1 /{\sqrt {3}}$, it can be shown that the potential is positive and
there are no instabilities in this sector either.
Therefore there are at least some hairy black holes which are stable under linear, spherically symmetric, perturbations.
It can further be proved that at least some of these solutions remain stable when non-spherically symmetric
perturbations are considered \cite{win:Sarbach1,win:ew3} but the analysis is highly involved and so we do not
attempt to summarize it here.

It should be remarked that it is unlikely that {\em {all}} nodeless black hole solutions are stable, although
this has not been investigated in the literature.
An in-depth study of the corresponding solitonic solutions \cite{win:Breitenlohner2} has revealed
that some soliton solutions for which $\omega (r)$ has no zeros, although they do not have any instabilities
in the sphaleronic sector, do possess unstable modes in the gravitational sector.
A scaling behaviour analysis of the solitonic solutions \cite{win:Hosotani1} has shown that the stable soliton solutions
can be approximated well by the stable solitons which exist on pure adS space.
On the other hand, the unstable solitons are interpreted as the unstable Bartnik-MacKinnon solitons \cite{win:Bartnik2}
dressed with solitons on pure adS.

\subsection{Other asymptotically anti-de Sitter ${\mathfrak {su}}(2)$ EYM solutions}
\label{win:sec:su2adSother}

\subsubsection{Dyonic solutions}
\label{win:sec:su2adSdyons}

In asymptotically adS, it is no longer the case that the only genuinely non-Abelian solutions
must have vanishing electric part in the gauge potential (\ref{win:eq:gaugepot}), so the
results of \cite{win:Ershov2,win:Ershov1} do not extend to non-asymptotically flat solutions.
As well as the magnetically charged solutions described above, dyonic black holes were
discussed in \cite{win:Bjoraker1,win:Bjoraker2}, which we shall not consider further here.
The stability of the dyonic solutions remains an open question as the perturbation equations
do not decouple into two sectors in this case, making analysis difficult.

\subsubsection{Topological black holes}
\label{win:sec:su2adStop}

As in Einstein-Maxwell theory, topological black hole solutions exist for ${\mathfrak {su}}(2)$ EYM in
adS \cite{win:Bij2}.
The metric in this case reads
\begin{equation}
ds^{2} = - \mu S^{2} \, dt^{2} + \mu ^{-1} \, dr^{2} +
r^{2} \, d\theta ^{2} + r^{2} f^{2}(\theta ) \, d\phi ^{2} ,
\end{equation}
where
\begin{equation}
f(\theta ) = \left\{
\begin{array}{lcl}
\sin \theta & & {\mbox {for $k=1$,}} \\
\theta & & {\mbox {for $k=0$,}}\\
\sinh \theta & & {\mbox {for $k=-1$,}}
\end{array}
\right.
\end{equation}
and
\begin{equation}
\mu = k - \frac {2m(r)}{r} - \frac {\Lambda r^{2}}{3}.
\end{equation}
The ansatz for the purely magnetic gauge field potential is now \cite{win:Bij2}
\begin{equation}
A = \tau _{x} \omega (r) \, d\theta + \left[ \tau _{y} \omega (r) + \tau _{z} \frac {d\ln f}{d\theta } \right]
f(\theta ) \, d\phi .
\end{equation}
When $\Lambda = 0$, only spherically symmetric solutions with $k=1$ are possible, but for $\Lambda <0$,
solutions with both $k=0$ and $k=-1$ have been found \cite{win:Bij2}.
All the solutions are nodeless, which can be easily proved from the field equations \cite{win:Bij2}.
It is found in \cite{win:Bij2} that all the $k=0$ solutions are stable under spherically symmetric perturbations in
both the sphaleronic and gravitational sectors.
The same is true for the $k=-1$ solutions for which $\omega >1$ as $r\rightarrow \infty $ \cite{win:Bij2}.

\subsubsection{Non-spherically symmetric solutions}
\label{win:sec:su2adSnonSS}

As in the asymptotically flat case, there are both soliton \cite{win:Radu7} and black hole \cite{win:Radu4}
solutions which are static but not spherically symmetric, so that the metric and gauge potential take
the form (\ref{win:eq:metriciso},\ref{win:eq:gaugeiso}).
Rotating black holes have also been found \cite{win:Mann2}, and there are also rotating dyonic soliton solutions
\cite{win:Radu5}.

\section{Asymptotically anti-de Sitter solutions for ${\mathfrak {su}}(N)$ EYM}
\label{win:sec:suNadS}

In the previous section we found that stable hairy black holes exist in ${\mathfrak {su}}(2)$ EYM
with a sufficiently large and negative cosmological constant.
A natural question is therefore whether there are stable hairy black hole solutions of
${\mathfrak {su}}(N)$ EYM in adS, and we examine this question in this section.

\subsection{Spherically symmetric numerical solutions}
\label{win:sec:num}

For any fixed $N$, the field equations (\ref{win:eq:YMe}) and (\ref{win:eq:Ee}) can be solved numerically
using standard techniques.
We will outline briefly some of the key features of the black hole solutions for ${\mathfrak {su}}(3)$ EYM.
Details of the corresponding soliton solutions and the solution space for ${\mathfrak {su}}(4)$ EYM can be found
in \cite{win:Baxter3}.

\begin{figure}
\includegraphics[angle=270,width=8cm]{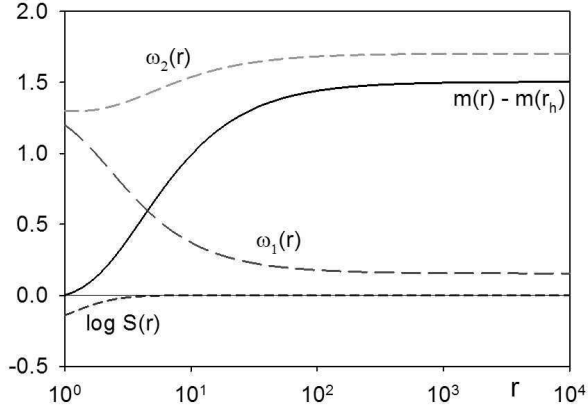}
\caption{Typical ${\mathfrak {su}}(3)$ black hole solution, with $r_{h}=1$, $\Lambda = -1$,
$\omega _{1}(r_{h}) = 1.2$ and $\omega _{2}(r_{h}) = 1.3$.
In this example, both gauge field functions have no zeros.
Taken from \cite{win:Baxter3}.}
\label{win:fig:su3bhex1}
\end{figure}

\begin{figure}
\includegraphics[angle=270,width=8cm]{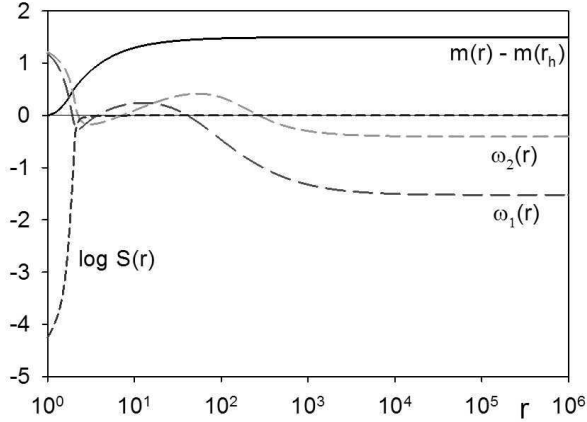}
\caption{Example of an ${\mathfrak {su}}(3)$ black hole solution, with $r_{h}=1$,
$\Lambda = -0.0001$, $\omega _{1}(r_{h})=1.184$ and $\omega _{2}(r_{h})=1.216$.
In this case, both gauge field functions have three zeros.
Taken from \cite{win:Baxter3}.}
\label{win:fig:su3bhex2}
\end{figure}

\begin{figure}
\includegraphics[angle=270,width=10cm]{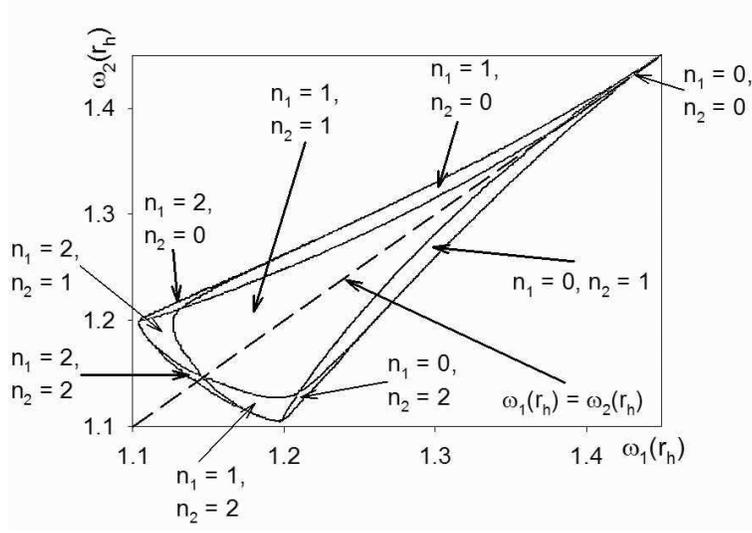}
\caption{Solution space for ${\mathfrak {su}}(3)$ black holes with $r_{h}=1$ and $\Lambda = -0.1$.
The numbers of zeros of the gauge field functions for the various regions of the solution space are shown.
For other values of $\omega _{1}(r_{h})$, $\omega _{2}(r_{h})$ we find no solutions.
There is a very small region containing solutions in which both gauge field functions have no zeros,
in the top-right-hand corner of the plot.
Taken from \cite{win:Baxter3}.}
\label{win:fig:su3bh3}
\end{figure}

\begin{figure}
\includegraphics[angle=270,width=9cm]{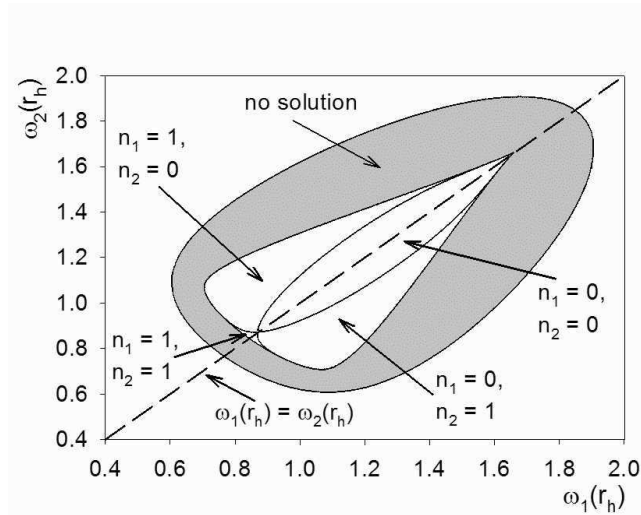}
\caption{Solution space for ${\mathfrak {su}}(3)$ black holes with $r_{h}=1$ and $\Lambda = -1$.
The shaded region indicates where the constraint (\ref{win:eq:su3bhconstraint}) is satisfied but we do
not find black hole solutions.
Outside the shaded region the constraint (\ref{win:eq:su3bhconstraint}) does not hold.
Where there are solutions, we have indicated the numbers of zeros of the gauge field functions
within the different regions.
For this value of $\Lambda $ there is a large region in which both gauge field functions have no zeros.
Taken from \cite{win:Baxter3}.}
\label{win:fig:su3bh2}
\end{figure}

For ${\mathfrak {su}}(3)$ EYM, there are two gauge field functions $\omega _{1}(r)$ and $\omega _{2}(r)$,
and therefore four parameters describing black hole solutions: $r_{h}$, $\Lambda $, $\omega _{1}(r_{h})$
and $\omega _{2}(r_{h})$.
Using the symmetry of the field equations (\ref{win:eq:omegaswap}),
we set $\omega _{1}(r_{h}),\omega _{2}(r_{h})>0$ without loss of generality.
The constraint (\ref{win:eq:constraint}) on the values of the gauge field functions at the horizon becomes, in this
case:
\begin{equation}
\left[ \omega _{1}(r_{h})^{2} - 2 \right] ^{2}
+ \left[ \omega _{1}(r_{h})^{2} - \omega _{2}(r_{h})^{2} \right] ^{2}
+ \left[ 2 - \omega _{2}(r_{h})^{2} \right] ^{2}
<
2 r_{h}^{2} \left( 1 - \Lambda r_{h}^{2} \right) .
\label{win:eq:su3bhconstraint}
\end{equation}
Two typical black hole solutions are shown in figures \ref{win:fig:su3bhex1} and \ref{win:fig:su3bhex2}.
The metric functions behave in a very similar way to the ${\mathfrak {su}}(2)$ solutions, smoothly interpolating
between their values at the horizon and at infinity.
We note that $S(r)$ in particular converges very rapidly to $1$ as $r\rightarrow \infty $.
In figure \ref{win:fig:su3bhex1}, we show an example of a black hole solution in which both gauge field functions
have no zeros.
We note that both gauge field functions are monotonic, however, one is monotonically increasing and the other
monotonically decreasing.
In our second example (figure \ref{win:fig:su3bhex2}) both gauge field functions have three zeros.
Although, in both our examples the two gauge field functions have the same number of zeros, we
also find solutions where the two gauge field functions have different numbers of zeros (see figures
\ref{win:fig:su3bh3} and \ref{win:fig:su3bh2}).

We now examine the space of black hole solutions.
Since we have four parameters, in order to produce two-dimensional figures, we need to fix two parameters in each case.
We find that varying the event horizon radius produces similar behaviour to the ${\mathfrak {su}}(2)$ case, so
for the remainder of this section we fix $r_{h}=1$ and consider the phase space for different, fixed values of $\Lambda $,
scanning all values of $\omega _{1}(r_{h})$, $\omega _{2}(r_{h})$ such that the constraint (\ref{win:eq:su3bhconstraint})
is satisfied.
From the discussion in section \ref{win:sec:suN}, we have embedded ${\mathfrak {su}}(2)$
black hole solutions when, from (\ref{win:eq:embeddedsu2}):
\begin{equation}
\omega _{1}(r)={\sqrt {2}}\omega (r)=\omega _{2}(r)
\end{equation}
which occurs when $\omega _{1}(r_{h})=\omega _{2}(r_{h})$.

\begin{figure}
\includegraphics[angle=270,width=9cm]{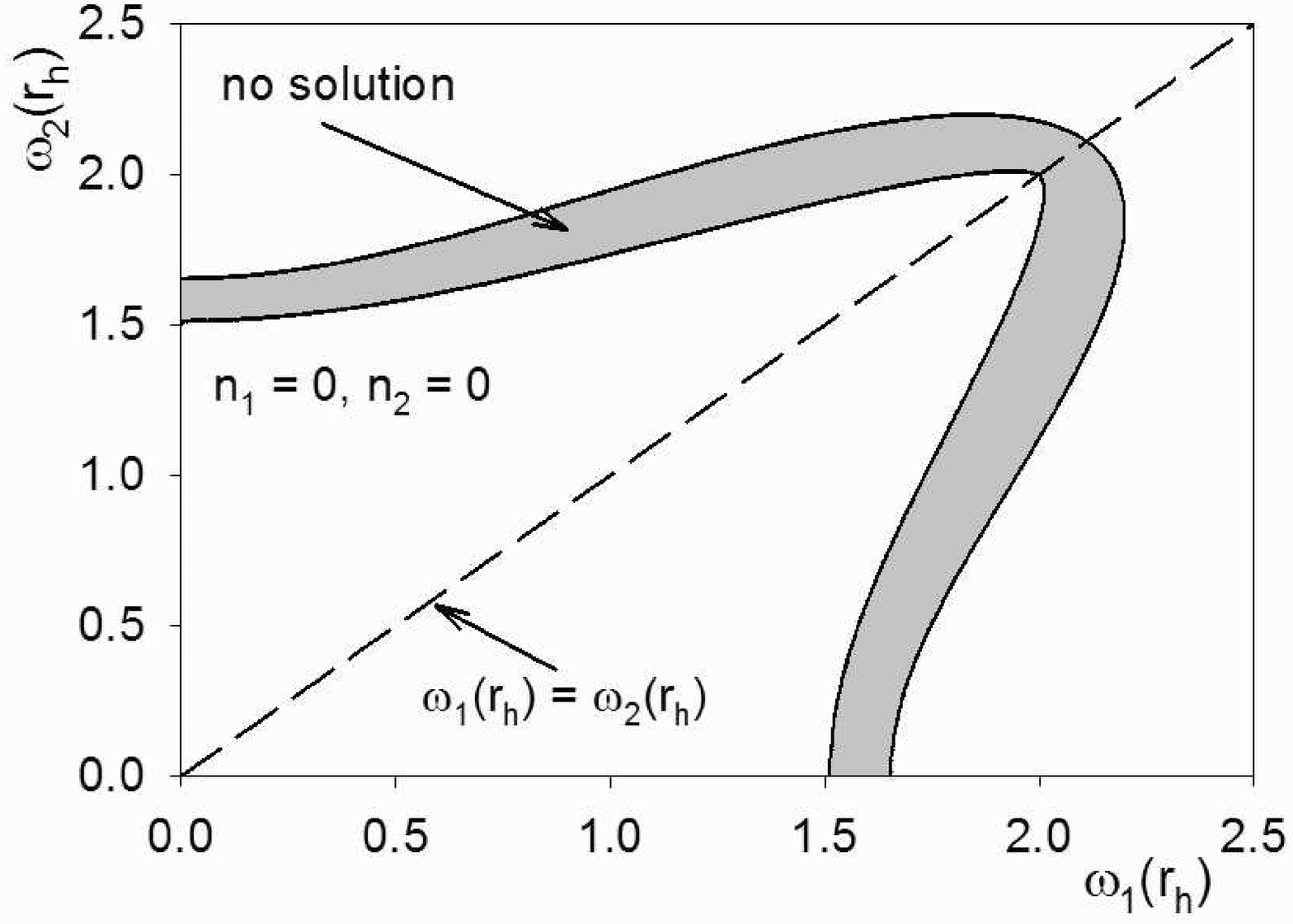}
\caption{Solution space for ${\mathfrak {su}}(3)$ black holes with $r_{h}=1$ and $\Lambda = -5$.
It can be seen that for the vast majority of the phase space for which the constraint (\ref{win:eq:su3bhconstraint})
is satisfied, we have black hole solutions in which both gauge field functions have no zeros.
Taken from \cite{win:Baxter3}.}
\label{win:fig:su3bh4}
\end{figure}

In figures \ref{win:fig:su3bh3}-\ref{win:fig:su3bh4} we plot the phase space of solutions for fixed
event horizon radius $r_{h}=1$ and varying cosmological constant $\Lambda = -0.1$, $-1$ and $-5$ respectively.
In each of figures \ref{win:fig:su3bh3}-\ref{win:fig:su3bh4} we plot
the dashed line $\omega _{1}(r_{h})=\omega _{2}(r_{h})$,
along which lie the embedded ${\mathfrak {su}}(2)$ black holes.
It is seen in all these figures that the solution space is symmetric about this line, as would be expected from
the symmetry (\ref{win:eq:Nswap}) of the field equations.
The solution space is found to be symmetric about the line $\omega _{1}(r_{h})=\omega _{2}(r_{h})$ not only in terms of
where we find solutions, but also in terms of the numbers of zeros of the gauge field functions.
To state this precisely, suppose that at the point $\omega _{1}(r_{h})=a_{1}$, $\omega _{2}(r_{h})=a_{2}$
we find a black hole solution in which $\omega _{1}(r)$ has $n_{1}$ zeros and $\omega _{2}(r)$ has $n_{2}$ zeros.
Then, at the point $\omega _{1}(r) = a_{2}$, $\omega _{2}(r) = a_{1}$, we find a black hole solution
in which $\omega _{1}(r)$ has $n_{2}$ zeros and $\omega _{1}(r)$ has $n_{1}$ zeros.
This is clearly seen in figures \ref{win:fig:su3bh3} and \ref{win:fig:su3bh2}, and follows from the symmetry
(\ref{win:eq:Nswap}) of the field equations.
As we increase $\left| \Lambda \right| $, we find
(see figures \ref{win:fig:su3bh3}-\ref{win:fig:su3bh4})
that the solution space expands as a proportion of the
space of values of $\omega _{1}(r_{h})$, $\omega _{2}(r_{h})$ satisfying the constraint (\ref{win:eq:su3bhconstraint}).
It can also be seen from figures \ref{win:fig:su3bh3}-\ref{win:fig:su3bh4} that the number of nodes of the gauge
field functions decreases as $\left| \Lambda \right| $ increases, and that the space of solutions becomes simpler.
For $\Lambda = -0.1$, there is a very small region of the solution space where both gauge field functions have
no zeros.
This region expands as we increase $\left| \Lambda \right| $, until for $\Lambda =-5$, both gauge field
functions have no zeros for all the solutions we find.

The solution space becomes progressively more complicated as $N$ increases, due to the increased number of parameters
required to describe the solutions.
However, the key feature described above is found; namely that for sufficiently large $\left| \Lambda \right| $, all
the solutions we find are such that all the gauge field functions $\omega _{j}$ have no zeros.
These solutions are of particular interest since one might hope that at least some of them might be stable.

\begin{figure}
\includegraphics[width=10cm]{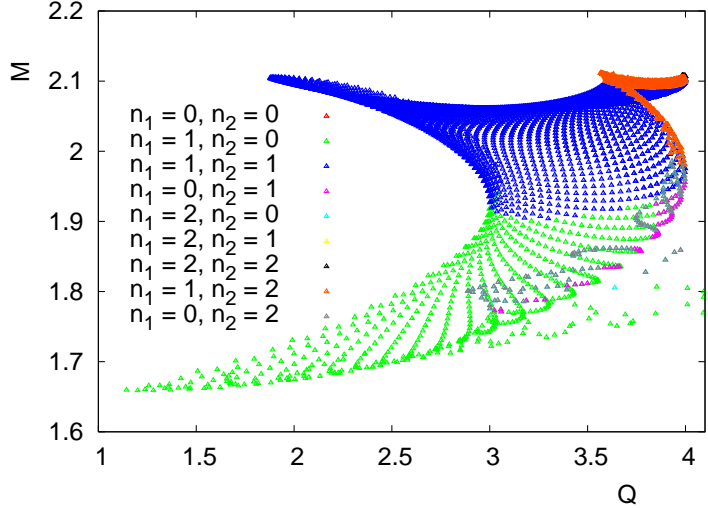}
\caption{Black hole mass $M$ versus magnetic charge $Q$ for ${\mathfrak {su}}(3)$ EYM black holes with
$r_{h}=1$ and $\Lambda = -0.1$.
There are many different combinations of numbers of zeros of the gauge field functions (see figure \ref{win:fig:su3bh3}),
which are indicated by different colours.
Here we have performed a scan over a grid of possible values of the gauge field functions at the event horizon,
$\omega _{1}(r_{h})$, $\omega _{2}(r_{h})$, leading to discrete points in the spectrum.
This is to enable the complicated structure of the spectrum to be seen.}
\label{win:fig:su3charge}
\end{figure}

As with the ${\mathfrak {su}}(2)$ black holes we may consider the spectra of
black hole solutions by plotting the relationship between the mass $M$ and magnetic charge $Q$ of the
solutions (see figure \ref{win:fig:su2charge1} for the ${\mathfrak {su}}(2)$ case).
As may be expected, for higher $N$ the spectra are even more complicated than for ${\mathfrak {su}}(2)$.
In figure \ref{win:fig:su3charge} we plot some of the possible values of $M$ and $Q$ for ${\mathfrak {su}}(3)$
EYM black holes with $\Lambda = -0.1$ and $r_{h}=1$.
In figure \ref{win:fig:su3charge} we have colour-coded the various possible numbers of zeros of the gauge field
functions (cf. figure \ref{win:fig:su3bh3}).
We have used a discrete grid of initial values of the gauge field functions at the event horizon
$(\omega _{1}(r_{h}),\omega _{2}(r_{h}))$ and plotted discrete points so that at least some of the structure can be seen.
In this case, because we have a four-parameter $(\Lambda , r_{h}, \omega _{1}(r_{h}), \omega _{2}(r_{h}))$
space of solutions of the field equations, even when $\Lambda $ and $r_{h}$ are fixed, we obtain two-dimensional
regions in the $(M,Q)$ plane, rather than curves as in the ${\mathfrak {su}}(2)$ case.
It can be seen from figure \ref{win:fig:su3charge} that the spectrum is very complicated, with the regions
corresponding to different numbers of zeros of the gauge field functions overlapping.
It is certainly the case that the black holes cannot be uniquely characterized by the four parameters
$\left( \Lambda, r_{h}, M, Q \right) $.

\subsection{Analytic work}
\label{win:sec:analytic}

For any fixed value of $N$, it is possible to examine the space of solutions numerically.
However, we would like to know whether there are solutions for {\em {all}} $N$, and, in particular,
whether for all $N$ there are some solutions for which all the gauge field functions have no zeros,
which we expect to be the case for sufficiently large $\left| \Lambda \right| $.
Answering this question for general $N$ requires analytic rather than numerical work.

In \cite{win:ew1}, the existence of black hole solutions for which the gauge function $\omega (r)$ had no zeros was
proven analytically in the ${\mathfrak {su}}(2)$ case.
Since ${\mathfrak {su}}(2)$ solutions can be embedded as ${\mathfrak {su}}(N)$ solutions via (\ref{win:eq:embeddedsu2}),
we have automatically an analytic proof of the existence of nodeless ${\mathfrak {su}}(N)$ EYM black holes in adS.
However, these embedded solutions are ``trivial'' in the sense that they are described by just three parameters:
$r_{h}$, $\Lambda $ and $\omega ( r_{h} ) $.
The question is therefore whether the existence of ``non-trivial'' (that is, genuinely ${\mathfrak {su}}(N)$)
solutions in which all the gauge field functions $\omega _{j}(r)$ have no zeros can be proven analytically.
The answer to this question is affirmative, and involves a generalization to ${\mathfrak {su}}(N)$ of the
continuity-type argument used in \cite{win:ew1}.
The details are lengthy and will be presented elsewhere \cite{win:Baxter1}.
Here we simply outline the key steps in the proof.

\begin{figure}
\includegraphics[angle=270,width=6cm]{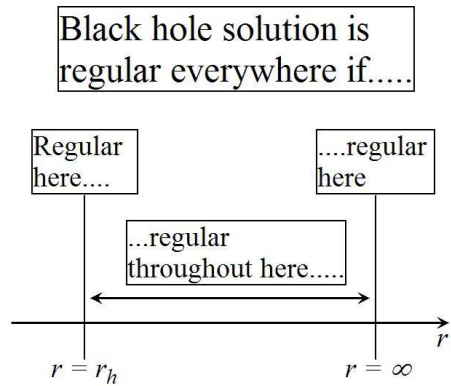}
\caption{Sketch of the main steps in the proof of the existence of non-trivial
${\mathfrak {su}}(N)$ EYM black holes in adS for which all the gauge field functions have no zeros.
We wish to find black hole solutions which are regular on the event horizon, regular everywhere outside
the event horizon, and regular at infinity.
We thank J.~E.~Baxter for providing this sketch.}
\label{win:fig:flow}
\end{figure}

The main idea of the proof is sketched in figure \ref{win:fig:flow}.
We wish to find black hole solutions which are regular on the event horizon, regular everywhere outside
the event horizon, and regular at infinity.
The proof proceeds via the following steps:
\begin{enumerate}
\item
We firstly prove (generalizing the analysis of \cite{win:Kunzle2} to include $\Lambda $)
that the field equations (\ref{win:eq:YMe},\ref{win:eq:Ee}) and initial conditions at the event horizon
(\ref{win:eq:horizon})
possess, locally in a neighborhood of the horizon, solutions which are analytic in
$r$, $r_{h}$, $\Lambda $ and the parameters $\omega _{j} ( r_{h} ) $.
As might be expected, the analysis of \cite{win:Kunzle2} requires only minor modifications to include
a negative cosmological constant.
\item
This enables us to prove that, in a sufficiently small neighborhood of any embedded ${\mathfrak {su}}(2)$
solution in which $\omega (r)$ has no nodes, there exists (at least in a neighborhood of the
event horizon) an ${\mathfrak {su}}(N)$ solution in which all the $\omega _{j}(r)$ have
no nodes.
\item
Using the analyticity properties of the solutions of the field equations, we then show that
these ${\mathfrak {su}}(N)$ solutions can be extended out to large $r_{L}\gg r_{h}$, provided the initial parameters
$\omega _{j} ( r_{h} ) $ are sufficiently close to those of an embedded ${\mathfrak {su}}(2)$ solution
in which $\omega (r)$ has no zeros.
Furthermore, by analyticity, none of the $\omega _{j}(r)$ will have any zeros between
the event horizon $r_{h}$ and $r_{L}$.
\item
The key part of the proof lies in then showing that these ${\mathfrak {su}}(N)$ solutions
can be further extended out to $r\rightarrow \infty $ and that they
satisfy the boundary conditions (\ref{win:eq:infinity}) at infinity.
This part of the analysis uses the properties of the Yang-Mills field equations (\ref{win:eq:YMe})
in the asymptotically adS regime.
As in the ${\mathfrak {su}}(2)$ case \cite{win:ew1}, these have very different properties from the
asymptotically flat case, and this makes it much easier to prove the existence of solutions.
Furthermore, it can be shown that the gauge field functions $\omega _{j}(r)$ will have no zeros
for $r\ge r_{L}$.
\end{enumerate}
In summary, this process gives genuinely ${\mathfrak {su}}(N)$ black hole solutions in which all the gauge
field functions have no zeros, and which are characterized by the $N+1$ parameters
$r_{h}$, $\Lambda $ and $\omega _{j} ( r_{h} ) $.

\subsection{Stability analysis of the spherically symmetric solutions}
\label{win:sec:stab}

The remaining outstanding question is whether these new black holes,
with potentially unbounded amounts of gauge field hair, are stable.
We consider linear, spherically symmetric perturbations only for simplicity.
The analysis of \cite{win:Sarbach1,win:ew3} in the ${\mathfrak {su}}(2)$ case
revealed that, for sufficiently large $\left| \Lambda \right| $, stability under spherically symmetric perturbations
continued to hold also for non-spherically symmetric perturbations, and one might hope that a similar result will
hold in the more complex ${\mathfrak {su}}(N)$ case.
However, we leave this for future work.
Even for spherically symmetric perturbations, the analysis is highly involved in the ${\mathfrak {su}}(N)$ case
and the details will be presented elsewhere \cite{win:Baxter2,win:Baxter1}.
Here we briefly outline just the key features.
The perturbation equations themselves can be found in section \ref{win:sec:pert}.

\subsubsection{Sphaleronic sector}
\label{win:sec:sphaleronicstability}

\begin{figure}
\includegraphics[angle=270,width=9cm]{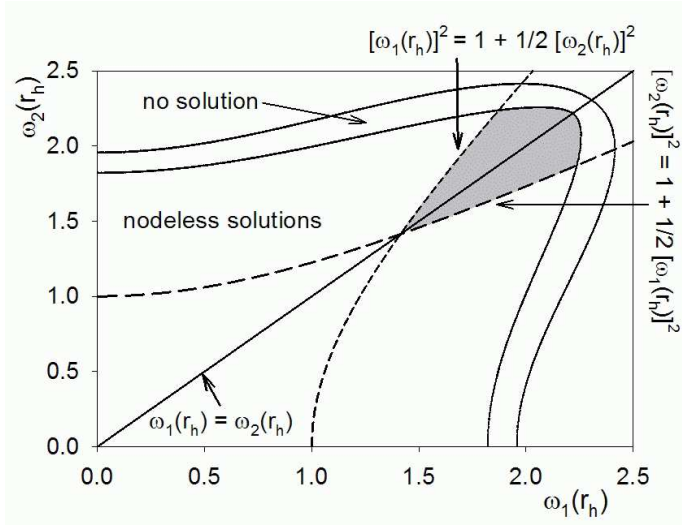}
\caption{Phase space of black hole solutions in ${\mathfrak {su}}(3)$ EYM with $\Lambda = -10$ and $r_{h}=1$.
The shaded region shows where solutions exist which satisfy the inequalities (\ref{win:eq:stabineq}) at the
event horizon.
Taken from \cite{win:Baxter4}.}
\label{win:fig:su3stabregion}
\end{figure}

The sphaleronic sector consists of the perturbation equations (\ref{win:eq:deltabeta},\ref{win:eq:deltaPhi})
together with the Gauss constraint (\ref{win:eq:GC}).
The analysis of this sector essentially follows that of \cite{win:Brodbeck1} in the asymptotically flat case.
We begin by defining yet more new variables, $\delta \epsilon _{j}$, for $j=1,\ldots ,N$ by
\begin{equation}
\delta \epsilon _{j} = r {\sqrt {\mu }} \delta \beta _{j} ,
\end{equation}
then, after much algebra, the sphaleronic sector perturbation equations
can be cast in the form
\begin{equation}
-{\ddot {{\underline {\Psi }}}} = {\cal {M}}_{S}{\underline {\Psi }},
\end{equation}
where the $(2N-1)$-dimensional vector ${\underline {\Psi }}$ is defined by
\begin{equation}
{\underline {\Psi }} = \left(  \delta \epsilon _{1}, \ldots , \delta \epsilon _{N}, \delta \Phi _{1}, \ldots ,
\delta \Phi _{N-1} \right) .
\end{equation}
and ${\cal {M}}_{S}$ is a self-adjoint, second order, differential operator (involving derivatives with respect to $r$
but not $t$), depending on the equilibrium functions $\omega _{j}(r)$, $m(r)$ and $S(r)$.
The operator ${\cal {M}}_{S}$ can be written as the sum of three parts.
The first is of the form $\chi ^{\dag }\chi $ for a particular first order differential operator $\chi $
(whose precise form can be found in \cite{win:Baxter2,win:Baxter1})
and is therefore manifestly positive and is regular if the gauge field functions $\omega _{j}$ have no zeros.
The second part vanishes when applied to a physical perturbation due to the Gauss constraint (\ref{win:eq:GC}).
The third part is a matrix ${\cal {V}}$ which does not contain any differential operators.
It can be shown that the matrix ${\cal {V}}$ is regular and positive definite provided the
unperturbed gauge functions $\omega _{j}(r)$ have no zeros and satisfy the $N-1$ inequalities
\begin{equation}
\omega _{j}^{2} >
1 + \frac {1}{2} \left( \omega _{j+1}^{2} + \omega _{j-1}^{2} \right)
\label{win:eq:stabineq}
\end{equation}
for all $j=1,\ldots N-1$, and all $r\ge r_{h}$.
The inequalities (\ref{win:eq:stabineq}) define a non-empty subset of the parameter space.
For example, we show in figure \ref{win:fig:su3stabregion} where the inequalities (\ref{win:eq:stabineq}) are satisfied
for the gauge field functions at the event horizon, for the particular case of $\Lambda = -10$ and $r_{h}=1$.
From figure \ref{win:fig:su3stabregion} we can see that there are some nodeless solutions which satisfy
the inequalities (\ref{win:eq:stabineq}) at the event horizon.
For any $N$, it can also be proved analytically that, for sufficiently large $\left| \Lambda \right| $, there
are non-trivial ${\mathfrak {su}}(N)$ solutions, in a neighbourhood of some embedded ${\mathfrak {su}}(2)$
solutions, such that the inequalities (\ref{win:eq:stabineq}) are satisfied at the event horizon.

\begin{figure}
\includegraphics[angle=270,width=9cm]{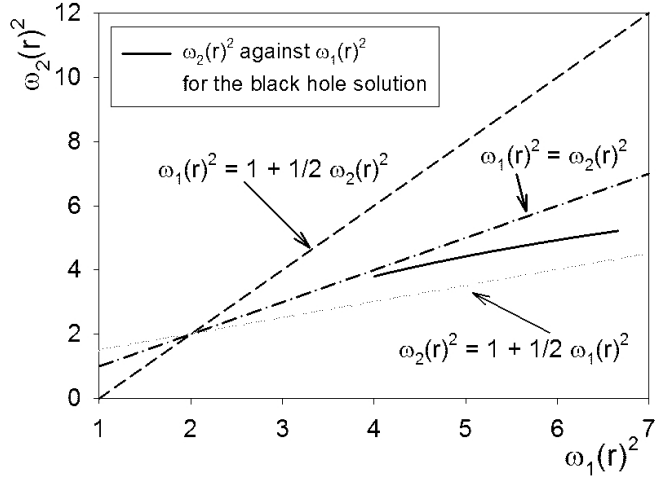}
\caption{An example of an ${\mathfrak {su}}(3)$ solution for which the inequalities (\ref{win:eq:stabineq})
are satisfied for all $r\ge r_{h}$.
In this example, $\Lambda = -10$, $r_{h}=1$ and the values of the gauge field functions at the event horizon
are $\omega _{1} ( r_{h} ) = 2$, $\omega _{2} ( r_{h} ) = 1.95$.
Taken from \cite{win:Baxter4}.}
\label{win:fig:su3stabex}
\end{figure}

However, the requirements of (\ref{win:eq:stabineq}) are considerably stronger, as the inequalities have to
be satisfied for {\em {all}} $r\ge r_{h}$.
Our analytic work shows that, in fact, for any $N$ and sufficiently large $\left| \Lambda \right| $, there
do exist solutions to the field equations for which the inequalities (\ref{win:eq:stabineq}) are indeed
satisfied for all $r$.
This involves proving that for at least some solutions for which the gauge field function values at the
event horizon lie within the region where the inequalities (\ref{win:eq:stabineq}) are satisfied,
the gauge field functions remain within this open region.
In figure \ref{win:fig:su3stabex} we show an example of such a solution for ${\mathfrak {su}}(3)$ EYM.

\subsubsection{Gravitational sector}
\label{win:sec:gravitationalstability}

As might be expected, the gravitational sector perturbation equations (\ref{win:eq:deltaomega}) are more
difficult to analyze than the sphaleronic sector perturbation equations.
For stable solutions, we require the matrix ${\cal {M}}_{G}$ (\ref{win:eq:calMG}) to be negative definite.
For sufficiently large $\left| \Lambda \right| $, it can be shown that ${\cal {M}}_{G}$ is indeed negative
definite for embedded ${\mathfrak {su}}(2)$ solutions, provided that $\omega ^{2}(r) > 1$ for all $r\ge r_{h}$
(the existence of such ${\mathfrak {su}}(2)$ solutions is proved, for sufficiently large $\left| \Lambda \right| $,
in \cite{win:ew1}).
As described in section \ref{win:sec:analytic} above,
our analytic work ensures the existence of genuinely ${\mathfrak {su}}(N)$
solutions in a sufficiently small neighborhood of these embedded ${\mathfrak {su}}(2)$ solutions.
These ${\mathfrak {su}}(N)$ solutions are such that the inequalities (\ref{win:eq:stabineq}) are satisfied for
all $r\ge r_{h}$ (and therefore the solutions are stable under sphaleronic perturbations).
The negativity of ${\cal {M}}_{G}$ can then be extended to these genuinely ${\mathfrak {su}}(N)$ solutions
using an analyticity argument, based on the nodal theorem of \cite{win:Amann1}
(see also \cite{win:ew3} for a similar argument for the non-spherically symmetric perturbations of the
${\mathfrak {su}}(2)$ EYM black holes).
The technical details of this argument will be presented elsewhere \cite{win:Baxter1}.

The conclusion of the work in this section is that there are at least some genuinely ${\mathfrak {su}}(N)$
EYM black holes in adS, for sufficiently large $\left| \Lambda \right| $,
for which all the gauge field functions $\omega _{j}$ have no zeros, and which
are stable under spherically symmetric perturbations in both the sphaleronic and gravitational sectors.

\section{Summary and outlook}
\label{win:sec:conc}

In this review we have studied classical, hairy black hole solutions of ${\mathfrak {su}}(N)$ EYM theory, particularly
spherically symmetric space-times and black holes in adS.
We very briefly discussed some of the key aspects of the solutions in asymptotically flat space, which
have been extensively reviewed in \cite{win:Volkov1}.
Hairy black hole solutions exist for all $N$, with $N-1$ gauge field degrees of freedom \cite{win:Mavromatos2},
however, all these solutions are unstable \cite{win:Brodbeck1}.
Therefore, while these hairy black holes violate the ``letter'' of the no-hair conjecture (that is, their geometry
is not completely fixed by global charges measurable at infinity), its ``spirit'' is maintained.
In particular, stable equilibrium black holes are comparatively simple objects, described completely by just a few
parameters.

The main conclusion of this article is that this is not true in adS.
The existence of stable hairy black holes in ${\mathfrak {su}}(2)$ EYM \cite{win:ew1} did not really contradict the
``spirit'' of the no-hair conjecture, as only a single additional parameter was required to fix the geometry
outside the event horizon.
However, the recent work \cite{win:Baxter4} which shows that there are stable hairy black holes in
${\mathfrak {su}}(N)$ EYM in adS for arbitrarily large $N$ changes the picture completely.
For sufficiently large $\left| \Lambda \right| $, an infinite number of parameters are required in order
to describe stable black holes.
We might flippantly describe these as ``furry'' black holes, since they possess copious amounts of hair.

What are the consequences for black hole physics in adS of these ``furry'' black holes?
These need to be explored.
Given the huge amount of interest in the adS/CFT correspondence in string theory
\cite{win:Maldacena1,win:Witten2,win:Witten3}, a natural question is how black hole hair
in the bulk asymptotically adS space-time relates to the dual CFT.
In particular, it has been suggested \cite{win:Hertog1} that there should be observables in the dual (deformed) CFT
which are sensitive to the presence of black hole hair.
Another example of this approach can be found in \cite{win:Gauntlett1}, where an adS/CFT interpretation is given
of some stable seven-dimensional black holes with ${\mathfrak {so}}(5)$ gauge fields.
We would expect that, in analogy with the ${\mathfrak {su}}(2)$ case
\cite{win:Chamseddine1,win:Chamseddine2,win:Gubser1,win:Hubscher1,win:Mann2,win:Radu6,win:Radu8},
there are solutions in some super-gravity theories with a gauge group containing an ${\mathfrak {su}}(N)$ factor,
which will need to be studied in the context of adS/CFT.
There is evidence \cite{win:Mavromatos1} that there are non-trivial black hole solutions of ${\mathfrak {su}}(\infty )$
EYM in adS, giving black holes not just with unbounded amounts of hair, but infinite amounts of hair, at least
in the limit $\left| \Lambda \right| \rightarrow \infty $.
It remains to be seen whether exact solutions of the ${\mathfrak {su}}(\infty )$ field equations can be found for
finite $\Lambda <0$, and whether any of these black holes are stable.
If so, then their role in adS/CFT would be puzzling indeed.

Due to space restrictions, there are many aspects of black holes in EYM which we have not been able to discuss.
In particular, we have not mentioned the vast number of solutions which involve modifications of the EYM
action (\ref{win:eq:action}), including higher curvature terms
(see, for example, \cite{win:Kanti2,win:Kanti1})
or the inclusion of dilaton (see, for example, \cite{win:Radu3}),
Higgs (see, for example, \cite{win:Bij1,win:Lugo2,win:Lugo1})
or other modifications of the EYM action (see, for example, \cite{win:Moss1,win:Shiiki2,win:Shiiki1}).
Here we have also only studied four-dimensional space-times, while recent work has considered
EYM in higher-dimensional space-times (see, for example,
\cite{win:Breitenlohner4,win:Brihaye4,win:Brihaye3,win:Brihaye5,win:Brihaye6,win:Brihaye7,win:Brihaye1,win:Brihaye2,win:Hartmann1,win:Okuyama,win:Radu1,win:Radu2}
and \cite{win:Volkov7} for a review).

The black hole solutions of EYM and its variants certainly exhibit an abundantly rich structure, and no doubt
will have more surprises in store for us in the future.

\begin{acknowledgement}
We thank the organizers of the 4th Aegean Summer School for a most enjoyable and informative conference.
The work in section \ref{win:sec:suNadS} was done in collaboration with J.~E.~Baxter and Marc Helbling.
We thank B.~Bear, R.~F.~W.~Jackson and the Chester Chronicle for suggesting the terminology ``furry'' black holes.
We also thank Eugen Radu for numerous enlightening discussions.
This work was supported by STFC (UK), grant reference numbers PPA/G/S/2003/00082 and
PP/D000351/1.
\end{acknowledgement}

\end{document}